\documentclass[prb,twocolumn,showpacs,superscriptaddress,usenames]{revtex4}
\usepackage{bbm,bm}
\usepackage{amsmath,pstricks,pst-plot,natbib}
\usepackage{mathrsfs}
\usepackage{epsfig,psfrag}
\usepackage{amsfonts,amssymb}
\usepackage{graphicx}

\usepackage{dsfont}
\usepackage{psfrag}
\usepackage{dcolumn}

\definecolor{BrickRed}{cmyk}{0,0.89,0.94,0.28}
\definecolor{MidnightBlue}{cmyk}{0.98,0.13,0,0.43}
\definecolor{DarkGreen}{rgb}{0,0.7,0.1}

\newcount\minute
\newcount\hour
\newcount\hourMins
\def\now
{
  \minute=\time
  \hour=\time \divide \hour by 60
  \hourMins=\hour \multiply\hourMins by 60
  \advance\minute by -\hourMins
  \zeroPadTwo{\the\hour}:\zeroPadTwo{\the\minute}
}

\def\today
{
  \zeroPadTwo{\the\day}-\zeroPadTwo{\the\month}-\the\year%
}
\def\zeroPadTwo#1%
{
  \ifnum #1<10 0\fi
  #1%
}

\newcommand{\comment}[1] {}
\newcommand{\question}[1] {}

\begin{document}

\preprint{draft}

\newcommand{\tec}[1]{{\color{magenta} \{\small \sc #1\}}}
\newcommand{\tei}[1]{{\color{magenta} #1}}

\newcommand{\dif}{\mathrm{d}}
\newcommand{\bk}{{\mathbf k}}
\newcommand{\bx}{{\mathbf x}}
\newcommand{\bz}{{\mathbf z}}
\newcommand{\bq}{{\mathbf q}}
\newcommand{\bu}{{\mathbf u}}
\newcommand{\bn}{{\mathbf n}}
\newcommand{\bj}{{\mathbf j}}
\newcommand{\br}{{\mathbf r}}
\newcommand{\bR}{{\mathbf R}}
\newcommand{\bQ}{{\mathbf Q}}
\newcommand{\bN}{{\mathbf 0}}
\newcommand{\bK}{{\mathbf K}}
\newcommand{\bG}{{\mathbf G}}
\newcommand{\bX}{{\mathbf X}}
\newcommand{\bb}{{\mathbf b}}
\newcommand{\ba}{{\mathbf a}}
\newcommand{\bnabla}{{\bm \nabla}}
\newcommand{\tR}{\tilde R}
\newcommand{\tV}{\tilde V_L}

\newcommand{\mG}{\mathcal{G}}

\newcommand{\erf}{\mbox{erf}}
\newcommand{\erfc}{\mbox{erfc}}
\newcommand{\ds}{\displaystyle}
\newcommand{\cH}{{\cal H}}
\newcommand{\bphi}{\bm{\phi}}

\title{Pinning of Flux Lines by Planar Defects}

\author{Aleksandra Petkovi\' c}
\affiliation{Institut f\"ur Theoretische Physik, Universit\"at zu
K\"oln, Z\"ulpicher Stra\ss e 77, 50937 K\"oln, Germany}
\author{Thorsten Emig}
\affiliation{Institut f\"ur Theoretische Physik, Universit\"at zu
K\"oln, Z\"ulpicher Stra\ss e 77, 50937 K\"oln, Germany}
\affiliation{Laboratoire de Physique Th\'eorique et Mod\`eles
Statistiques, CNRS UMR 8626, Universit\'e Paris-Sud, 91405 Orsay,
France}
\author{Thomas Nattermann}
\affiliation{Institut f\"ur Theoretische Physik, Universit\"at zu
K\"oln, Z\"ulpicher Stra\ss e 77, 50937 K\"oln, Germany}

\date{\today\hspace{0.2cm} \now}

\begin{abstract}
  The influence of randomly distributed point impurities and planar
  defects on order and transport in type-II superconductors and
  related systems is studied. It is shown that the Bragg glass phase
  is unstable with respect to planar defects. Even a single weak
  defect plane oriented parallel to the magnetic field as well as to
  one of the main axis of the Abrikosov flux line lattice is a
  relevant perturbation in the Bragg glass. A defect that is aligned
  with the magnetic field restores the flux density oscillations which
  decay algebraically with the distance from the defect. The theory
  exhibits striking similarities to the physics of a Luttinger liquid
  with a frozen impurity. The exponent for the flux line creep in the
  direction perpendicular to a relevant defect is derived. We find
  that the flux line lattice exhibits in the presence of many randomly
  distributed parallel planar defects aligned to the magnetic field a
  new glassy phase which we call {\it planar glass}. The planar glass
  is characterized by diverging shear and tilt moduli, a transverse
  Meissner effect, resistance against shear deformations. We also obtain sample to
  sample fluctuations of the longitudinal magnetic susceptibility and
  an exponential decay of translational long range order in the
  direction perpendicular to the defects. The flux creep perpendicular
  to the defects leads to a nonlinear resistivity $\rho(J \to 0)\sim
  \exp[-(J_D/J)^{3/2}]$.  Strong planar defects enforce arrays of
  dislocations that are located at the defects with a Burgers vector
  parallel to the defects in order to relax shear strain.
\end{abstract}

\pacs{74.25.Qt, 71.55.Jv, 74.62.Dh,64.70.Rh}

\maketitle


\section{Introduction}
\label{sec:intro}

Type-I superconductors are both perfect conductors and perfect
diamagnets. In type-II superconductors the perfect diamagnetism is
reduced, an external magnetic field penetrates the sample above the
lower critical field $H_{c1}$ in the form of magnetic flux lines
(FLs)\cite{Tinkham}. A transport current will then lead to a motion of
the FLs, yielding a linear resistivity $\rho\approx\rho_nB/H_{c2}$ in
disorder-free samples. Here $B$ denotes the magnetic
induction and $H_{c2}$ is the upper critical
field\cite{Bardeen+65}. At $B=H_{c2}$ diamagnetism disappears
completely and $\rho$ reaches the resistivity $\rho_n$ of the normal
state. In order to recover the desired property of a
dissipation-free flow, FLs have to be pinned. Point defects, such as
vacancies or interstitials, are one type of pinning source. In
high-$T_c$ materials point impurities are almost always present due to
a non-stoichiometric composition of most materials. Impurity pinning
leads to a zero linear resistivity\cite{Blatter+94}. However, thermal
fluctuations allow for FL creep, resulting in a non-zero
\emph{nonlinear} resistivity of the form $\rho(J)\sim
\exp[-(J_P/J)^{\mu}]$ where the creep exponent is $\mu=1/2$ for point
impurities\cite{Nattermann90}. $J (\, \ll J_P)$ denotes the current
density and $J_P$ depends on $B$, temperature $T$, concentration and
strength of the pinning centers as well as on properties of the
material through the superconductor coherence length $\xi$ and the
penetration length $\lambda$. This FL creep law is closely related to
the order of the flux line lattice (FLL) in the presence of point
pinning centers.

The order of the FLL was a puzzle for a long time. Larkin concluded in
1970 that randomly distributed impurities destroy the long range order
of the Abrikosov lattice\cite{Larkin}.  Only much later it was
realized that the effect of impurities is weaker, resulting in a power
law decay of translational order of the FLs in the so-called ''Bragg
glass'' phase
\cite{Nattermann90,Giamarchi+94,Giamarchi+95,Emig+99,bogner+01,Klein+01}.

More effective pinning sources can further suppress the nonlinear
resistivity. One example are columnar defects, produced by heavy ion
radiation, that have been considered by Nelson and Vinokur
\cite{NeVi+92,NeVi+93}. These authors mapped the physics of FLs onto
the problem of the localization of bosons in two dimensions where FLs
play the role of world lines of the bosons\footnote{A similar approach
  was proposed also by \citet{Lyuksyutov92}}. At low temperatures they
found strongly localized FLs at the columnar defects, forming a ''Bose
glass'' phase. Thermally activated hopping of noninteracting FLs in
the limit $J\to 0$ leads to the creep exponent $\mu=1/3$, while FL
interactions yield the increased creep exponent $\mu=1$
\cite{Blatter+94}. The transport in this regime closely resembles the
variable range hopping of electrons in two dimensional disordered
semiconductors. This picture is expected to be valid for weak enough
applied magnetic fields, such that the density of defects is bigger
than the FL density. For a larger magnetic field, \citet{Radzihovsky95}
argued that the Bose glass coexists with a resistive liquid of
interstitial FLs which upon cooling freezes into a weakly pinned Bose
glass. For asymptotically weak currents, the creep of FL bundles
determines the nonlinear resistivity and $\mu=1$ is the creep
exponent\cite{Blatter+94}.
\begin{figure}
\includegraphics[width=0.95\linewidth]{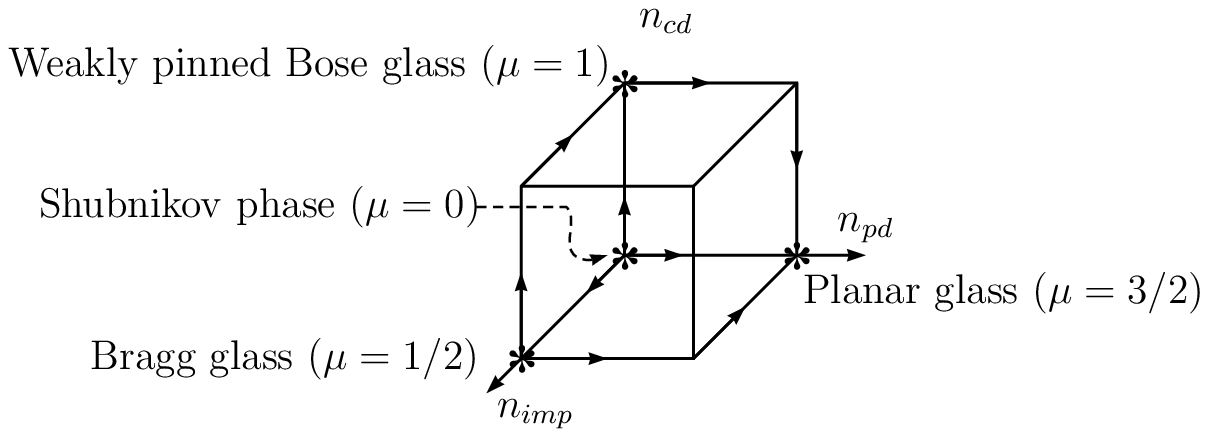}
\caption{Schematic phase diagram of disordered flux line lattices
    resulting from impurities, columnar and planar defects of
    concentration $n_{imp}$, $n_{cd}$ and $n_{pd}$, respectively. The
    stability of the phases with respect to different kinds of
    disorder is indicated by arrows.  $\mu$ denotes the creep
    exponent.}\label{Fig:1}
\end{figure}

In this work we consider planar defects like twin boundaries from
which even stronger pinning can be expected.  Twin boundaries are
ubiquitous in superconducting YBa$_2$Cu$_3$O$_{7-x}$ and La$_2$CuO$_4$
where they are needed to accommodate strains arising from tetragonal
to orthorhombic transition as a result of oxygen vacancy ordering and
due to rotation of the CuO$_{6}$ octahedra,
respectively\cite{many_orthogonal_twins}. Twin boundaries occur
frequently with the same orientation
\cite{Crabtree,Kwok+92,Oussena+96} or in orthogonal families of
lamellas (''colonies'')
\cite{many_orthogonal_twins,Sanfilippo+97}. They can be regularly
distributed with rather fixed spacing or with large variations in the
spacing\cite{Marchetti+95}. The mean distance $\ell_D$ of the defect
planes varies between $10\, nm$ \cite{Crabtree,Sanfilippo+97,Kwok+92}
and $1 \mu m$ \cite{Dolan,Oussena+96}. 

The common feature of all of the above mentioned defects is that they
lead to FL pinning, but what distinguishes them is the nature of the
pinned phase. In contrast to point disorder, which promotes FL
wandering, planar defects inhibit wandering and promote
localization. Pinning of individual FLs by columnar as well as planar
defects in the presence of bulk point disorder has been investigated
in the past\cite{Tang+93,BaKa+93,BaKa+94,Hwa+95}. The competition
between a planar defect and point impurities in three-dimensional
systems, for a single FL, leads to localization of the FL at all
temperatures\cite{Tang+93,BaKa+93,BaKa+94}. The influence of many
parallel defect planes on the creep of a single FL perpendicular to
the planes has been studied at low temperatures when the FL spacing
exceeds the average spacing between the planes\cite{Marchetti+95}.

The main focus of this paper is correlated disorder in the form of
planar defects. Some of the results of this paper have been published
earlier \cite{Emig+06,Petkovic+08}. Here we give additional
  results and present more detailed derivations. First, we discuss
the influence of a single planar defect on the stability of the Bragg
glass phase. Then we explore the effect of many defect planes on the
FLL. We find that the necessary condition for a planar defect to
become a relevant perturbation is that it is oriented parallel to the
magnetic field. In this case, its influence on the Bragg glass phase
can be characterized by the value of a single parameter
$g\equiv\frac{3}{8}\eta(a/\ell)^2$ which depends both on the exponent
$\eta$ describing the decay of the positional correlations in the
Bragg glass phase and on the orientation of the defect with respect to
the FLL. $a$ and $\ell$ are the mean spacing of the FLs in the absence of the defect plane and the distance between
lattice planes, of the Abrikosov lattice, parallel to the defect,
respectively. A weak defect turns out to be relevant if $g<1$, i.e.,
if it is parallel to one of the main crystallographic planes of the
FLL.

The FL density averaged over point impurities shows periodic order
with an amplitude decaying as a power law with the normal distance to
the defect plane. For a relevant defect on scales larger than $L_v$,
the exponent controlling the power law is $g$. For the definition of
$L_v$, see Eq.~(\ref{eq:Lv}) below. For $g>1$ a weak defect is
irrelevant (in the sense of renormalization group) and the density
profile decays faster, with a larger exponent $2g-1>1$, on scales
larger than the positional correlation length $L_a$. For a weak defect
tilted against the applied magnetic field we find that FL density
oscillations decay exponentially fast.  We investigate also the
dynamics of FL bundles perpendicular to the relevant defect for small
current densities $J\to 0$.  The nonlinear resistivity is $\rho(J)\sim
\exp{\left[-\frac{C_1}{J}\left(\log{\frac{C_2}{J}}\right)^2\right]}$
where $C_1$ and $C_2$ depend on
various parameters such as temperature, magnetic induction, density and strength of point
impurities as well as the strength of the defect plane.  We conclude
that a single relevant defect plane slows down the FL creep in
comparison to the BG phase.

There are interesting connections of some of the aspects of our
results to related two-dimensional classical or one-dimensional
quantum models.  A single planar defect in the Bragg glass phase
resembles the presence of a single columnar defect in a FLL confined
to a plane\cite{Hofstetter+04,Affleck+04,Radzihovsky} or a frozen
impurity in a Luttinger liquid\cite{Kane+92,Egger+95}. In all three
cases the bulk phases on both sides of the defect are characterized by
logarithmically diverging displacement correlations. The parameter $g$
plays the role of a temperature in the 2D classical case and of the
Luttinger liquid parameter in the 1D quantum case. The periodic order
seen around the defect plane has its counterpart in Friedel
oscillations around an impurity in Luttinger liquids.  Whereas in
the 1D (2D) case the relevance of an impurity is controlled
by tuning the interaction strength (temperature), in the present
case a change of $g$ can be accomplished by changing the orientation
of the defect with respect to the FLL. Transport properties of our
system are however different from the ones in the related systems.

In this article we also study the effect of a finite density of
randomly distributed parallel planar defects on the FLL at low
temperatures with the magnetic field aligned parallel to the
defects. We consider the case when the mean defect spacing is greater than FL spacing. Our results may be directly
applicable to a wide class of other systems with planar defects as a
stack of membranes under tension, charge density waves\cite{gruener},
domain walls in magnets and incommensurate systems\cite{BrCo78} since
we consider a simplified model with an uniaxial displacement field
perpendicular to the defects. We find a new phase, which we call
  {\it planar glass}, that is characterized by (i) diverging shear
(tilt) modulus that determines the energy cost for a shearing
{(tilting)} of the FLL in the direction perpendicular to the
planes; (ii) finite compressibility; (iii) sample to sample
fluctuations of the longitudinal magnetic susceptibility; (iv)
an exponential decay of positional correlations in the direction
perpendicular to the defects and (v) a creep exponent $\mu=3/2$ for
creep in the direction perpendicular to the defects for small currents
$J\to 0$. The planar glass is different from the Bragg glass or the Bose glass phase
and from the phase found for equally spaced defects \cite{BaNe94}. The planar glass is
stable over a finite range of tipping angles of the applied
  magnetic field away from the direction parallel to the planar
defects, i.e., it is characterized by a transverse Meissner effect. Similarly, the planar glass is characterized by a resistance against shear deformations that are perpendicular to the defects. Naturally, realistic
samples contain both point and correlated disorder (as columnar and/or
planar defects). We find that the planar glass is stable against both
weak point and weak columnar disorder. The schematic phase diagram is
shown in Fig.~\ref{Fig:1}.  When we consider a {\it vector}
displacement field, we additionally find that strong disorder enforces
arrays of dislocations in order to relax shear strain. They are
located at the defects with a Burgers vector parallel to the defects.
We argue that the main properties of the planar
glass remain unchanged by dislocations.

The paper is organized as follows. In Sec.~\ref{section:BG} we
introduce a model for interacting FLs that couple to weak point
impurities and briefly review known results of this model. In
Sec.~\ref{section:single} we consider a single defect plane as a
perturbation to the Bragg glass phase and study the FL order using the
a renormalization group analysis. A finite density of randomly
distributed defects is explored in Sec.~\ref{section:weak} and the
novel planar glass is characterized. Functional renormalization group
equations are derived in $d=6-\epsilon$ dimensions. The response to
tilt and shear deformations is discussed as well as sample to sample
fluctuations of the longitudinal magnetic susceptibility. The
positional correlation functions are computed and the stability of
planar glass with respect to point and columnar disorder is
studied. In Sec.~\ref{section:strong} we consider the limit of {\it
  strong} planar defect potentials. In Sec.~\ref{sec:Creep} we
consider the FL dynamics for small currents by investigating FL creep
in the presence of a single defect plane, both with and without point impurities, and in the presence of a finite density of planes. Finally, in
Sec.~\ref{section:conclusions} we discuss a model with a vector
displacement field. Technical details and list of recurrent symbols are relegated to the Appendices.

\section{The Bragg Glass phase}
\label{section:BG}

In this section we summarize some known results on pinning effects due
to randomly distributed point impurities for interacting FLs. We use
elasticity theory to describe a dislocation free array of FLs (for a
review see, e.g., Blatter et al.\cite{Blatter+94}). Undistorted FLs
are exactly parallel to the $z$-axis which we assume to be the direction of the
applied magnetic field.  The FLs form a triangular Abrikosov FLL
in the $xy$-plane with a lattice constant $a$. In
order to describe distortions of the FLs from the perfect lattice
positions $\bR_{\nu}$, we use a two-component vector displacement
field $\bu_{\nu}(z)$. Since we are interested
in the behavior on large length scales, it is appropriate to describe
the interacting FLs in terms of a continuum elastic approximation with
a continuous displacement field $\bu_{\nu}(z)\to \bu(\br)$. 

The Hamiltonian
\begin{align}\label{eq:Hamiltonian}
\cH=\cH_0+\cH_{P}
\end{align}
consists of the elastic energy of the FLL,
$\cH_0$, and pinning energy of point impurities, $\cH_{P}$. The elastic
energy of the dislocations free FLL reads
\begin{align}\label{eq:elastic Ham}
\cH_{0}=\frac{1}{2}\int
\frac{d^2\bq_{\perp}dq_z}{(2\pi)^3} \widetilde{\bu}(\bq)\left(\widetilde{\mathcal{G}}_L^{-1}\mathbf{P}_L+\widetilde{\mathcal{G}}_T^{-1}\mathbf{P}_T\right)
  \widetilde{\bu}(-\bq),
\end{align}
where $\bq_{\perp}=q_x\hat{\mathbf{x}}+q_y\hat{\mathbf{y}}$. $\mathbf{P}_L^{ij}=q_{i}q_{j}/q_{\perp}^2$ and
$\mathbf{P}_T^{ij}=\delta_{ij}-q_{i}q_{j}/q_{\perp}^2$
are projectors onto the longitudinal and transversal modes,
respectively, with propagators
\begin{align}
\widetilde{\mathcal{G}}_L^{-1}=c_{11}\bq_{\perp}^2+c_{44}q_z^2,\\
\widetilde{\mathcal{G}}_T^{-1}=c_{66}\bq_{\perp}^2+c_{44}q_z^2.
\end{align}
In general, the compression ($c_{11}$) and the tilt ($c_{44}$) moduli
are non-local on length scales smaller than the penetration depth
$\lambda$ but the shear modulus $c_{66}$ is always local.  However,
the dispersion of $c_{11}$ and $c_{44}$ on small length scales is
negligible for the present problem, since we are interested in
asymptotic properties at large length scales and small currents.
Hence, in the following we introduce a cutoff in momentum space given
by $\Lambda\approx {2\pi}/{\lambda}$ and neglect the non-locality of
the elastic moduli. The elastic Hamiltonian of Eq.~(\ref{eq:elastic
  Ham}) can be obtained from symmetry arguments\cite{Landau7}. The ideal FLL is isotropic in $xy-$plane and has
$D_{6h}$ symmetry group.


The pinning energy of randomly distributed point impurities is modeled by the coupling
\begin{align}\label{eq:pointpinning}
\cH_{P}=\int d^3\br\;\rho(\br,\bu)V_{P}(\br),
\end{align}
of the local FL density
$\rho(\br,\bu)=\sum_{\nu}\delta(\bx-\bR_{\nu}-\bu_{\nu}(z))$ to the
pinning potential $V_{P}(\br)$, where $\bx=(x,y)$. From this definition and the Poisson summation
formula\cite{Emig+99,Giamarchi+95} the density can be also written as
\begin{align}
\label{eq:density FLs}\rho\left(\br,\bu(\br)\right)= \rho_0+\rho_0\left\{- \bm{\nabla}_{\bx} \bu(\br) +
\sum_{\bG\neq 0}e^{i\bG[\bx-\bu(\br)]} \right\}
\end{align}
where $\rho_0=2/({\sqrt
3}a^2)$, and $\bG$ is a vector of
 the reciprocal lattice. 
$V_{P}=-v_p\sum_{i}\delta_{\xi}(\bx-\bx_i)\delta(z-z_i)$ represents the pining potential due to randomly distributed point impurities. The $\delta$-functions are considered to have a finite width of the order of the superconductor coherence length $\xi$. For simplicity we subtract the average of the random potential and look at fluctuations around the average value. The pinning potential then satisfies
\begin{align}\label{eq: correpoint}
\overline{V_{P}(\br)}=0, \quad
\overline{V_{P}(\br)V_{P}(\br')}=n_{imp}v_p^2\delta_{\xi}(\bx-\bx')\delta(z-z').
\end{align}
The strength of the disorder is characterized by $v_{p}^2n_{imp}$.
Higher order correlations of the (unrenormalized) pinning potential are nonzero but for weak disorder can be neglected. The restriction to two-point correlations
of $V_P$ leads to the same replica Hamiltonian one obtains when $V_P$ would be Gaussian distributed.

The model given by Eqs.~(\ref{eq:Hamiltonian})--(\ref{eq: correpoint})
has been studied in detail using perturbation theory\cite{Larkin},
Flory--type arguments\cite{Feigelman+89,Nattermann90}, a Gaussian
variational ansatz\cite{Korshunov93,Giamarchi+94,Giamarchi+95} and
functional renormalization group
\cite{Giamarchi+94,Giamarchi+95,Emig+99,bogner+01}.  The correlations
of the FLL fluctuations change with length scale and are characterized
by three different regimes: the Larkin or random force regime (RF),
the random manifold regime (RM) and the Bragg glass (BG) phase. These
regimes are distinguished by the scaling behavior of
\begin{align}\label{eq:roughness}
\overline{\langle(
\bu(\br)-\bu(\mathbf{0}))^2\rangle}\propto|\br|^{2\zeta}\, ,
\end{align}
which defines the roughness exponent $\zeta$. Here $\langle
\ldots\rangle$ denotes a thermal and $\overline{\cdots}$ a disorder
average.

(i) In the Larkin regime \cite{Larkin} the displacements are
sufficiently small so that the FLs stay within one minimum of the
disorder potential $V_{P}(\br)$ and perturbation theory can be
applied. The effect of the disorder potential on the FLL is properly
described by a random force $F_{P}(\bR_{\nu},z)=-\bnabla_{\bx}
V_{P}(\bR_{\nu},z)$. The roughness exponent is $\zeta_{RF}=(4-d)/{2}$,
where $d$ denotes dimension of the system, so that the positional
correlation function
\begin{align}\label{eq:positional correlation}
S_{\bG}(\br)=\overline{\langle e^{i \bG \bu(\br)}e^{-i \bG \bu(\mathbf{0})}\rangle}
\end{align}
decays exponentially fast in $d=3$. The Fourier transform of $S_{\bf
  G}({\bf r})$ is the structure factor which can be directly measured
in diffraction experiments. The Larkin lengths $L_{\xi}^z$ and
$L_{\xi}^{x}$ are defined as the crossover length scales where the
conditions $\overline{\langle
  (\bu(\mathbf{0},z=L_{\xi}^z)-\bu(\mathbf{0}))^2\rangle}\propto\xi^{2}$
and $\overline{\langle
  (\bu(|\bx|=L_{\xi}^x,0)-\bu(\mathbf{0}))^2\rangle}\propto\xi^{2}$
are satisfied. This leads to
\begin{align}\label{eq: Larkinlength}
L_{\xi}^z&\simeq\frac{\phi_0\xi^6}{Bv_{p}^2n_{imp}}\frac{c_{44}c_{66}}
{1+\kappa}\notag\\
L_{\xi}^x&\simeq\frac{\phi_0\xi^6}{Bv_{p}^2n_{imp}}
\frac{c_{44}^{1/2}{c_{66}}^{3/2}}{1+\kappa^{3/2}} \, ,
\end{align}
where $\phi_0=h c/(2 e)=2.07\;10^{-7} G\;cm^2$ is the flux quantum and
$\kappa=c_{66}/c_{11}$.  The length $L_{\xi}$ increases with
decreasing disorder strength. An increase in the magnetic
induction $B$ effectively increases the disorder strength so that
$L_{\xi}$ shrinks.

(ii) On scales greater than the Larkin length, a description in terms
of random forces become inapplicable and the RM regime applies. In the
RM regime FLs explore many minima of the disorder potential but the
typical displacement is still smaller than the FL spacing $a$. Hence
the FLs do not compete with neighboring FL for identical pinning
centers.  A Flory--type argument\cite{Kardar87, Nattermann87} yields
the roughness exponent $\zeta_{RM}=(4-d)/{6}$ but in
Ref.~[\onlinecite{Emig+99}] is was shown within an $\epsilon=4-d$
expansion that $\zeta_{RM}$ depends on the ratio
$\kappa={c_{66}}/{c_{11}}$ and varies between $0.1737\epsilon$ and
$0.1763\epsilon$. The positional order decays according to a stretched
exponential,
\begin{align}
S_{\bG}(\br)\sim \exp{\left[-\frac{G^2 r^{2\zeta_{RM}}}{2}\right]} \, .
\end{align}

(iii) On length scales larger than the positional correlation length
$L_a$ the RM regime becomes inapplicable. $L_{a}^{z,\bx}\approx
L_{\xi}^{z,\bx}\left({a}/{\xi}\right)^{{1}/{\zeta_{RM}}} $ is defined
as the scale at which the mean square displacement of FLs is of the
order $a$. Therefore, it is crucial to keep the periodicity
$\bu\rightarrow\bu+\bR_{\nu}$ of the interaction between FLs and point
disorder\cite{Nattermann90}. This leads to a much slower {\it logarithmic}
increase ($\zeta_{BG}=0$) of the elastic displacement of the FLs than in
the Larkin and RM regime. It was shown
\cite{Nattermann90,Korshunov93,Giamarchi+94,Giamarchi+95,Emig+99,bogner+01}
that thermal fluctuations are irrelevant and that the pinned FLL
exhibits a power law decay of positional correlations,
\begin{align}\label{eq:eta}
S_{\bG}(\bx,0)\sim |\bx|^{-\eta_{\bG}},
\end{align}
where $\eta_{\bG}=\eta(G/G_0)^2$ and $G_0=4\pi /(\sqrt{3} a)$.  This
result resembles the correlations in {\it pure} 2D crystals at finite
temperatures.  A functional renormalization group analysis in
$d=4-\epsilon$ dimensions yields a non-universal exponent $\eta$ that
varies with the elastic constants of the FLL
\cite{Emig+99,bogner+01}. Extrapolating to $d=3$, one finds only a
very weak variation with $1.143<\eta<1.159$
\cite{Emig+99,bogner+01}. Despite of the glassy nature of the phase
algebraically divergent Bragg peaks still exist which motivated the
name Bragg glass \cite{Giamarchi+94,Giamarchi+95}. The existence of
the Bragg glass phase has been experimentally confirmed
\cite{Klein+01,PRLbragg}. 

After this summary of the scaling regimes, we briefly review  the
replica theory for the Hamiltonian of
Eq.~(\ref{eq:Hamiltonian}). Using the replica method, we average over
point impurities (see Appendix A) and obtain the replica
Hamiltonian 
\begin{align}\label{eq:replica H}
\cH^n_P=\sum_{\alpha=1}^n\cH_0(\bu^{\alpha})-\frac{1}{2T}\sum_{\alpha,\beta=1}^n\int
 d^3\br \; R_P[\bu^{\alpha}(\br)-\bu^{\beta}(\br)].
\end{align}
\begin{align}\label{eq:R(u)}
R_P(\bu)=(v_p\rho_0)^2 n_{imp}\sum_{\bG\neq0} e^{i\bG\bu}\delta_{\xi^{-1}}(\bG),
\end{align}
where $\delta_{\xi^{-1}}(\bG)$ is the delta function smeared out over
a region of size $\xi^{-1}$.  The correlation functions $C_T$ and
$C_D$ that describe thermal fluctuations and disorder induced
fluctuations can be written as
\begin{align}\label{eq:correlations}
C_T&=\overline{\langle \widetilde{\bu}(\bq)\widetilde{\bu}(-\bq)\rangle-\langle \widetilde{\bu}(\bq)\rangle\langle \widetilde{\bu}(-\bq)\rangle}\notag\\
&=(2\pi)^d T\big\{\widetilde{\mathcal{G}}_L(\bq)
+\widetilde{\mathcal{G}}_{T}(\bq)\big\}\notag\\
C_D&=\overline{\langle \widetilde{\bu}(\bq) \rangle\langle\widetilde{\bu}(-\bq)\rangle}=(2\pi)^d\Delta(\bq) \big\{\widetilde{\mathcal{G}}_L^2(\bq) +\widetilde{\mathcal{G}}_{T}^2(\bq)\big\} \, ,
\end{align}
where the last equation defines $\Delta(\bq)$, which we shell obtain in harmonic approximation below.

In the next section we study the interplay between point impurities
and a planar defect. This is a difficult problem since we have to deal
with two nonlinear terms.  We consider the planar defect as a
perturbation to the BG fixed point and examine the stability of the BG
phase. Also we explore the effects of the defect on the order of the
FLL. In the following we will use an effective quadratic Hamiltonian
that reproduces the displacement correlations of
Eq.~(\ref{eq:roughness}) of the full nonlinear disordered model
Eq.~(\ref{eq:replica H}). A systematic analysis must be based on
$\epsilon=4-d$ expansion, and a functional renormalization group
analysis shows that displacements obey Gaussian statistics to lowest
order in $\epsilon$\cite{Wiese}. It should be noted that the effective
Hamiltonian does not capture all physics, in particular, it cannot
describe correctly the FL dynamics since it cannot reproduce the
energy barriers for FL motion\cite{Hwa+94}. An effective
quadratic Hamiltonian has been also used for a model with an uniaxial
displacement to study a dislocation mediated transition of the
FLL\cite{Kierfeld+2000}.

The effective quadratic replica Hamiltonian in $d$ dimensions reads\cite{Emig+99,bogner+01}
\begin{align}\label{eq:Replica Hamiltonian quadratic}
\cH^n_0=\frac{1}{2}\sum_{\alpha,\beta=1}^{n}(2\pi)^{-d}\int d^d\bq\;
\widetilde{\bu}^{\alpha}(\bq)\;\widetilde{\boldsymbol{\mG}}^{-1}_{\alpha,\beta}(\bq)\;\widetilde{\bu}^{\beta}(-\bq)
\end{align} where
\begin{align}\label{eq:propagator}
\widetilde{\boldsymbol{\mG}}^{-1}_{\alpha,\beta}(\bq)=&\delta_{\alpha,\beta}
\Big(\widetilde{\mathcal{G}}_L^{-1}(\bq)\mathbf{P}_L+\widetilde{\mathcal{G}}_T^{-1}(\bq)\mathbf{P}_T
\notag\\&+n\frac{\Delta(\bq)}{T}\mathds{1}\Big)-
\frac{\Delta(\bq)}{T}\mathds{1}.
\end{align}
It yields the correlation functions of Eq.~(\ref{eq:correlations}),
where $\Delta(\bq)$ describes the behavior of
$\Delta=-\partial^2_{u_x}R_P(\mathbf{0})=-\partial^2_{u_y}R_P(\mathbf{0})$
on different length scales. Using a functional renormalization
group in $d=4-\epsilon$ dimensions, it has been shown that to lowest order
in $\epsilon$\cite{Emig+99,bogner+01}
\begin{align}\label{eq:correlator}
\Delta(\bq)\sim \left\{ \begin{array}{ll}
1, & \frac{1}{L_{\xi}}\lesssim q \lesssim \Lambda\\
q^{\epsilon-2\zeta_{RM}}, & \frac{1}{L_a}\lesssim q \lesssim \frac{1}{L_{\xi}}\\
\epsilon q^{\epsilon}, & q \lesssim \frac{1}{L_a}\, .
\end{array} \right.
\end{align}
The function $\Delta(\bq)$ reaches the fixed point form $q^{\epsilon}\Delta^*(\kappa) c_{44} c_{66} a^2 $ in the BG
phase, where $\Delta^*(\kappa)\sim \epsilon /(1+\kappa)$ depends only
on elastic constants but not on the disorder strength. We note that
Emig et al.\cite{Emig+99,bogner+01} have obtained their results by
calculating the integrals, needed for the RG equations, systematically
for $d=4$ with a two-dimensional vector $\bz$. The results are then
extended to three dimensions by setting $\epsilon=1$ in
$\Delta^*(\kappa)$. This approach does not influence the main physics
(like the logarithmic roughness of FL in the BG phase), but may
influence the dependence of exponents $\eta$ and $\zeta_{RM}$ on
the elastic constants. In this way the dimensionality of $z$ ''axis'' and
the contribution of the term $c_{44}\bq_z^2$ in the propagators are
more weighted than the other axes and other terms $\sim
c_{66},c_{11}$, respectively. In the following, in order to not
overestimate the effect of a planar defect that is parallel to the $z$
axis, we will calculate all the integrals in $d=3$ if not stated
otherwise. If the numerical values of $\eta$ and $\zeta_{RM}$ are
important for our conclusions, we will comment on a possible influence
that the use of results found by Emig et al.\cite{Emig+99,bogner+01}
can have.

\section{Single defect}
\label{section:single}

In this section the influence of a \emph{single} planar defect on the
Bragg glass order of the FLL is studied. In some parts of this
section, when examining FLs density oscillations around the defect, we
will study the isotropic limit with $c_{11}=c_{44}=c_{66}=c$ in order
to focus on the important physics. In this limit the propagators read
$\widetilde{\mathcal{G}}_L^{-1}(\bq)=\widetilde{\mathcal{G}}_T^{-1}(\bq)=c
q^2$. By this assumption, only the weak dependence of $\eta$ and
$\zeta_{RM}$ on the elastic constants is ignored.

\subsection{Model}

The pinning energy of a planar defect can be written in the form
\begin{align}\label{eq:defectHamiltonian}
\cH_{D}=\int d^3\br\rho(\br,\bu)V_{D}(\br\cdot\bn_D-\delta) \, ,
\end{align}
where $V_D(\br\cdot\bn_D-\delta)$ is the potential of the defect
plane.  $\bn_D$ and $\delta$ denote the unit vector perpendicular to
the defect plane and its distance (along $\bn_D$) from the origin of
the coordinate system, respectively. The Bragg glass order that we are
interested in is dominated by disorder fluctuations on large length
scales where microscopic details become irrelevant. Therefore we may
approximate the defect potential by a smeared out $\delta$-function,
$V_{D}(x)\approx -v\delta_{\xi}(x)$. Since the superconducting order
is reduced in the defect plane, it is plausible to assume $v>0$ (for
more details, see Section IX of \citet{Blatter+94}). When we assume
that FLs gain condensation energy when they overlap with the
defect plane, a rough estimate for the defect strength is $v\approx
H_{c}^2\xi^3$ with $H_c$ the thermodynamic critical field.

\begin{figure}
\includegraphics[width=0.5\linewidth]{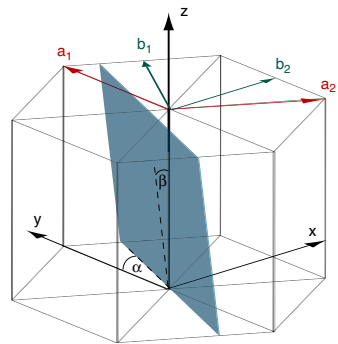}
\caption{Vectors of the triangular flux line lattice ($\ba_1$,
  $\ba_2$) and of its reciprocal lattice ($\bb_1$, $\bb_2$), and the
  angles $\alpha$, $\beta$ that define the orientation of the defect
  plane.}
\label{Fig:breaks}
\end{figure}

In order to integrate over the delta function of the defect potential,
it is convenient to introduce an explicit parametrization for the
position vector $\br_D$ of the defect plane which obeys $\br_D\cdot
\bn_D=\delta$. With the parametrization
\begin{eqnarray}
\label{eq:defect-para} \br_D&=&(\bx_{D},z_D)+\delta
\bn_D,\,\,\,z_D=t \cos \beta\\\nonumber \bx_{D}&=&(s \sin \alpha -
t\cos\alpha\sin\beta, s\cos\alpha+t\sin\alpha\sin\beta)\\ \nonumber
\bn_D&=&(\cos{\beta}\cos{\alpha},-\cos{\beta}\sin{\alpha},\sin{\beta})
\end{eqnarray}
we introduce in-plane coordinates $s$, $t$, and the two angles
$\alpha$ and $\beta$ which determine the rotation of the plane
with respect to the $y$- and $z$-axis, respectively (see Fig.
  \ref{Fig:breaks}).
The defect energy now reads
\begin{align}
\label{eq:energy_defect_2}
\cH_D=v\rho_0\!  \int \!  dt\,ds\,dr_{\perp}\delta_{\xi}(r_{\perp}-\delta)
\Big\{\bm{\nabla}_{\bx} \bu(t,s,r_{\perp}) \notag \\ -\sum_{\bG\ne \bN}
e^{i\bG[r_{\perp}\bn_D+\bx_{D}-\bu(t,s,r_{\perp})]}\Big\} \, ,
\end{align}
where $r_{\perp}=\br\cdot\bn_D$.  Since the displacement field $\bu$
varies slowly on the scale of the FLL constant $a$, the integrals over
$s$ and $t$ vanish for all $\bG$ with the exception of those for which
the oscillatory factor $e^{i\bG \bx_{D}}$ is unity (for all $s$, $t$).
This condition can be satisfied only if $\sin\beta=0$, i.e., if the
defect plane is \emph{parallel} to the applied magnetic field. There
remains a second condition for the angle $\alpha$ which results from
the constraint that $\bG=m \bb_1+n \bb_2$, with integer $m$ and $n$,
has to be perpendicular to $\bx_{D}$.  Expressing the defect plane
(for $\sin\beta=0$) as $\bx_{D}=(c_1\ba_1-c_2\ba_2)s$ where
$\ba_i\bb_j=2\pi\delta_{ij}$, one sees that the second condition is
equivalent to the condition $m/n=c_2/c_1$. Hence if $c_1/c_2$ is
irrational, the effect of the defect plane is always averaged to
zero. On the other hand, for rational $c_2/c_1$ we may choose
$m_D,n_D$ to be the smallest coprime pair with $c_2/c_1=m_D/n_D$.
Then $m_D$, $n_D$ are the Miller indices of the defect plane and only
those $\bG$ which are integer multiples of $\bG_D=m_D \bb_1+n_D \bb_2$
contribute in Eq.~(\ref{eq:energy_defect_2}).  In the following, we
will concentrate on the contribution from these $\bG$-vectors
only. The FLL planes (of the ideal lattice) that are parallel to a
defect plane with Miller indices $m_D$, $n_D$ have a separation of
$\ell=\frac{\sqrt 3}{2}a/\sqrt{m_D^2+m_Dn_D+n_D^2}$ and hence
$G_D=2\pi/\ell$.

For the defect plane aligned to the magnetic field we take the
$x$-axis to be perpendicular to the defect (i.e.~$\alpha=\beta=0$) and
hence the defect Hamiltonian becomes
\begin{align}\label{eq:defect H}
\cH_D= \rho_0v\!\int\!
dy dz\Big\{\bnabla_{\bx}\bu(\br_{D})-\sum_{k>0}^{[{\ell}/{\xi}]_G}2
\cos{[kG_{D}(\delta-u_{x})]}\Big\},
\end{align}
where $\br_D=(\delta,y,z)$. Here $u_{x}$ denotes the component of the
displacement field that is perpendicular to the defect plane and
$[x]_G$ is the integer number that is closest to $x$.

\subsection{Renormalization group analysis}

In this subsection we discuss the influence of the planar defect on
the stability of the BG phase using a renormalization group (RG)
analysis. We employ a sharp-cutoff scheme by integrating out the
displacement field $\widetilde{\bu}^>(\bq)$, with wave vectors $\bq$
in an infinitesimal momentum shell below the cutoff
$\Lambda>|\bq|>\Lambda/b=\Lambda e^{-l}$ and subsequently rescale
lengths and momenta according to
\begin{align}
\bq'&=\bq b\\
\br'&=\frac{\br}{b}.
\end{align}
We split the displacement field into weakly varying modes $u^{<}(\br)$
and strongly varying modes $u^>(\br)$ that include Fourier components
out of and in the momentum shell, respectively. We choose to not
rescale the field $\bu'({\br'})=\bu^<(\br)$ which implies a
rescaling of its Fourier transform, 
$\widetilde{\bu}'(\bq')={\widetilde{\bu}^<(\bq)}/{b^3}$.

The defect plane is considered as a perturbation to the Hamiltonian of
Eq.~(\ref{eq:Replica Hamiltonian quadratic}).  The gradient term of
Eq.~(\ref{eq:defect H}) scales $\sim L$ if the defect size $\sim
L^2$. Since the elastic energy Eq.~(\ref{eq:elastic Ham}) scales in
the same way, the gradient term is a marginal perturbation. It can be
also eliminated by the transformation
$u'_x(\br)=u_x(\br)+\frac{v\rho_0}{2c_{11}}\;\mathrm{sgn}(x-\delta)$,
where $\mathrm{sgn}(0)=0$. This transformation does neither change the
terms $\sim c_{66}$, $c_{44}$ of Eq.~(\ref{eq:elastic Ham}) nor the
pinning energy due to point impurities in Eq.~(\ref{eq:replica H})
since all replica fields are transformed in the same way. The gradient
term of the defect pinning energy tends to increase the FL density at
the defect as can be seen from the transformation above.

In order to account for different renormalization of the harmonic
components of the defect pinning energy, we introduce the variables
$v_k$ for the strengths of the harmonics of order $k$. A cumulant
expansion yields to first order in $v$ the renormalization
\begin{align}\label{eq:flow}
\frac{v_k(l)}{T(l)}&=\frac{v}{T}e^{2l}\langle
\cos{[kG_Du_{x}^{\alpha,>}(\br_D)]}\rangle \notag\\ &=\frac{v}{T}e^{2l}
e^{-\frac{1}{2} (k G_D)^2 \langle
[u_{x}^{\alpha,>}(\br_D)]^2\rangle} \notag\\&=\frac{v}{T}e^{(2-k^2 g)l},
\quad g=\frac{3}{8}\eta\left(\frac{a}{\ell}\right)^2 \, ,
\end{align}
where the factor $e^{2l}$ is due to a rescaling of lengths.  $\langle
[u_{x}^{\alpha,>}(\br_D)]^2\rangle$ is obtained at the BG fixed point
to linear order in $l$. Due to the irrelevance of thermal
fluctuations, we have neglected contributions that come from the
thermal part of the propagator of Eq.~(\ref{eq:propagator}).  We have
chosen to rescale temperature instead of elastic constants in order to
organize the RG analysis of the zero temperature BG fixed point. It is
important to note that Eq.~(\ref{eq:flow}) holds only on length scales
larger than the positional correlation length $L_a$.

In the random manifold regime $\overline{\langle
  e^{i{\bG}\bu(\br)}\rangle}$ decays with the system size as a
stretched exponential and the effect of the defect plane is reduced by
disorder fluctuations on intermediate length scales. Hence, the
renormalized and rescaled value of the defect strength is reduced to
\begin{align}\label{eq:v_k}
v_k\approx v(L_a/a)^2 e^{-\mathcal{C}(G_Dka)^2}
\end{align}
on the scale $L=L_a$, where $\mathcal{C}$ is a positive constant. This value is
the initial defect strength $v_k(l=0)$ to be used in
Eq.~(\ref{eq:flow}).

The RG flow equations in the Bragg glass regime now read
\begin{align}\label{eq:RGequations}
\frac{dT}{dl}&=-T\\\label{eq:RG-defect-flow}
\frac{dv_k}{dl}&=v_k(1-k^2 g).
\end{align}
Hence $v_1$ is a relevant perturbation provided $g<1$, i.e., if
\begin{equation}
\label{eq:relevance} \eta(m_D^2+m_Dn_D+n_D^2)<2\,\, \text{or}
\quad \ell
> \sqrt{\frac{3\eta}{8}}\,\,a\approx 0.66 \,\,a,
\end{equation}
which is compatible only with $\ell={\sqrt 3}a/2\approx 0.87 a$.  A
relevant defect plane must be oriented parallel to one of the three
main crystallographic planes of the FLL (i.e.~$\cos2\beta=\cos
6\alpha= 1$). When $\ell$ increases ($g$ decreases) more FLs can gain
energy from the defect plane and hence render it more relevant.

\begin{figure}
\includegraphics[angle=90,width=0.5\linewidth]{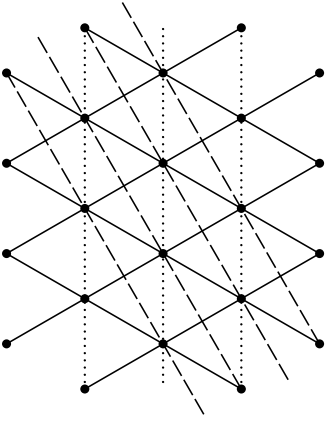}
\caption{Two possible orientations of defect planes relative to the
  flux line lattice, corresponding to $g=\eta/2$ (dotted lines) and
  $3\eta/2$ (dashed lines).}
\label{Fig:orientation}
\end{figure}

Emig et al.~\cite{Emig+99,bogner+01} have calculated $\eta$ in a
one-loop functional RG expansion in $4-\epsilon$ dimensions.  Higher
loops as well as the fact that all integrals are evaluated in $d=4$
with $\bz$ being a two dimensional vector, may influence the actual
numerical value of the coefficient $\eta$ in $d=3$. However, we argue
that this higher order correction does not affect our conclusion that
a single defect is relevant only if it is parallel to the main
crystallographic planes, since $g= \eta(m_D^2+m_Dn_D+n_D^2)/2$ can
change only in finite steps ($\eta/2$, $3\eta/2$, $7\eta/2$, $\dots$)
when rotating the defect plane (see Fig.~\ref{Fig:orientation}). 
\subsection{Effective Hamiltonian}

In this section we discuss whether higher order cumulants in $v$ can
lead to a renormalization of the parameter $g$ and hence can influence
the condition for the relevance of a defect that was derived in the
previous section.  The renormalization described by
Eq.~\eqref{eq:RG-defect-flow} does not occur in the bulk but on the
defect plane. Hence it is possible to develop an effective theory that
is defined on the defect plane only. Since the defect couples only to
the displacement $u_x$ on the defect plane, we integrate out $u_x$
outside the defect and $u_y$ across the entire sample. This
integration is facilitated by employing the effective Gaussian
  theory for the BG phase of the previous Section. At $T=0$ we are interested in the
ground state and hence we solve the Euler--Lagrange equation for
$\bu(\br)$ with the condition $u_{x}(\br_D)=\varphi(\br_D)$ at the
defect plane, where $\varphi(\br_D)$ is an arbitrary function. An
  equivalent functional integral approach is presented in Appendix
  \ref{app:effective}. The effective replica Hamiltonian on the
defect plane reads
\begin{align}\label{eq:H_2D}
\cH_{eff}^n=-\sum_{\alpha}\sum_{k>0} 2 v_k \rho_0 \int d\br_D\;
\cos{\{kG_D[\delta-\varphi^{\alpha}(\br_D)]\}}
\notag \\+\frac{1}{2}\sum_{\alpha,\beta} \frac{1}{(2\pi)^{d-1}}\int
\;d^{d-1}\;\bq\; \widetilde{\varphi}^{\alpha}(\bq)\;
\widetilde{\boldsymbol{\mathcal{Q}}}^{-1}_{\alpha,
\beta}(\bq)\;\widetilde{\varphi}^{\beta}(-\bq),
\end{align}
where $\bq$ is the in-plane momentum,
\begin{align}\label{eq:Q}
\langle \widetilde{u}_{x}^{\alpha}(x,\bq) \widetilde{u}_{x}^{\beta}(x,\bq') \rangle =T(2\pi)^d\delta(\bq+\bq')\widetilde{\boldsymbol{\mathcal{Q}}}^{\alpha,\beta}(\bq)
\end{align}
and $v_k$ are renormalized parameters on the scale of the positional
correlation length.  In order to avoid technical complications, we
consider the limit of isotropic elasticity. In line with an $\epsilon$
expansion, we evaluate the integrals in $d=4$ and then set
$\epsilon=1$ in the expression for fixed point value
$\Delta^*(\kappa)$. This approach does not affect our conclusion and
leads to a clearer result. On scales larger than $L_a$ we get
\begin{align}
\mathcal{Q}^{-1}_{\alpha,\beta}(\bq)=&2\sqrt{\frac{c n \Delta_{BG} }{T}+(qc)^2}\delta_{\alpha,\beta}\notag\\&-\frac{2c\Delta_{BG}}{T}\left(cq+\sqrt{\frac{c n \Delta_{BG}}{T}+(qc)^2}\right)^{-1}\, ,
\end{align}
where $c$ is the elastic constant and $\Delta_{BG}=c^2 a^2 \Lambda
\Delta^*(1)$. The same procedure can be performed also in the RF and
RM regimes using the corresponding quadratic Hamiltonian in $d$
dimensions. The effective Hamiltonian in $d-1$ dimensions has a long
ranged elasticity (term $\sim q$ in the limit $\Delta_{BG}\to 0$) that
results from the local bulk elasticity. A RG analysis of
the effective Hamiltonian of Eq.~(\ref{eq:H_2D}) shows that neither
$\Delta_{BG}$ nor the elastic constants are renormalized, and hence
$g$ is not renormalized. From this we conclude that a weak defect is a
relevant perturbation only for $g<1$.


\subsection{Density oscillations}

Next, we study the order of the FLs next to a defect plane. We
consider separately the case of a relevant and an irrelevant planar
defect plane.  For simplicity, we assume isotropic elasticity and
choose to place the origin of the coordinate system on the defect
plane, i.e., wet set $\delta=0$. In the absence of a planar defect,
FL density fluctuations due to point impurities obey
$\overline{\langle \rho(\br)-\rho_0\rangle}=0$. The defect plane pins
FLs and yields a {\it long-ranged} restoration of the translational
order parameter $e^{i \bG\bu(\br)}$. We find Friedel--like
oscillations of the FL density with an amplitude that decays as a
power law with an exponent that depends on if the defect is relevant
or irrelevant in the RG sense.

\subsubsection{Irrelevant defect}\label{sec:fridelirellevant}

First, we consider an irrelevant defect parallel to the magnetic
field. The irrelevance of the defect potential for $g>1$ allows us to
compute the thermal and disorder average of FL density perturbatively
in the defect strength.
\begin{align}
\overline{\langle\delta\rho(\br,\bu(\br))\rangle}=\lim_{n\to 0}{\prod_{\alpha=1}^n}\int \mathcal{D}\bu^{\alpha}\; \delta\rho(\br,\bu^{\gamma}(\br))\; e^{-\beta\cH^n},
\end{align}
where $\delta\rho(\br,\bu)=\rho(\br,\bu)-\rho_0$, $\gamma$ is an
arbitrary replica index and
\begin{align}
\cH^n=\cH_0^n+\sum_{\alpha}\cH_D(\bu^{\alpha}) \, .
\end{align}
To lowest order in the defect strength we get
\begin{align}
\overline{\langle\delta\rho(\br,\bu(\br))\rangle}=\lim_{n\to 0}\langle\delta\rho(\br,\bu^{\gamma}(\br))\rangle=0 \, .
\end{align}
Even if the defect is irrelevant in the RG sense, it breaks the
translational symmetry perpendicular to the defect and hence modifies
the FL density locally. To correctly describe this effect we need to
compute the average change in the FL density to first order in $v$. We
find (see App.~\ref{app irrelevant})
\begin{align}
  \overline{\langle\delta\rho(\br,\bu(\br))\rangle}&=-\beta\lim_{n\to
    0}\sum_{\alpha=1}^n\langle\delta\rho(\br,\bu^{\gamma}(\br))\cH_{D}(\bu^{\alpha})\rangle.
\end{align}
In the limit $T\to 0$ this result can be expressed as
\begin{align}\label{eq:friedel irrelevant}
\overline{\langle\delta\rho(\br,\bu(\br))\rangle}&\approx\frac{v_1\rho^2_0G_D^2L_a}{c(2 g-1)} \cos{(G_D|x|)}\left(\frac{L_a}{|x|}\right)^{2 g-1}.
\end{align}
which becomes exact for large $x$. The result captures the large length scale
behavior for $L\ge L_a$. Here $v_1$ denotes the effective defect
strength on the scale $L_a$, cf.~Eq.~(\ref{eq:v_k}), and $|x|$ is the
normal distance from the defect plane. There are additional
contributions to Eq.~(\ref{eq:friedel irrelevant}) coming from the
higher harmonics in $\cH_D$. They are less important since they are
proportional to $v_k$ on the scale $L_a$ and they decay as
  $|x|^{-2 k^2 g+1}$ with $k\geq 2$. Although the defect is irrelevant
  in RG sense, it leads to Friedel--like oscillations in the density.

  If the defect plane is {\it not} parallel to the applied magnetic
  field, Friedel oscillations occur as well. However, the amplitude is
  exponentially suppressed. The amplitude of density oscillations with
  reciprocal lattice vectors $\bG=G(\cos{\alpha},-\sin{\alpha})$ (for
  a definition of the angles $\alpha$ and $\beta$ see
  Fig. \ref{Fig:breaks}) decays  beyond the
  distance $1/(G|\sin\beta|)$ from the defect plane. Similar physics
  occur in classical 2-dimensional systems with a columnar
  defect\cite{Hofstetter+04,Affleck+04}.
\subsubsection{Relevant defect}

The strength of a relevant defect grows under  renormalization
relative to the elastic and the impurity energy. On the scale
\begin{align}\label{eq:Lv} L_{v}\approx
  \max{\left\{L_a,L_a\left(\frac{c
          a^4}{vL_a}\right)^{1/(1-g)}\right\}}
\end{align}
the energies become of the same order and
perturbation theory breaks down. On larger scales, the defect
potential can be described effectively through the boundary condition
$u_{x}(x=0,y,z)=0$ for the displacement field at the defect
plane. With this constraint the system gains maximal energy from the
defect and the complete energy of the system is minimized. First, we
calculate displacement correlations
$\mG_{pin,ij}(x,x';\br_{\parallel}-\br'_{\parallel})=T^{-1}\overline{\langle
  u_{i}(\br)u_{j}(\br')\rangle}$ with the above boundary condition at the
defect and $\br=(x,\br_{\parallel})$.  We find in momentum space (see Appendix \ref{app relevant})
\begin{widetext}
\begin{align}\label{eq:correlation pinned Fourier}
\widetilde{\mG}_{pin,ij}(x,x;\bq)=\lim_{n\to 0}
 \left[  \widetilde{\boldsymbol{\mG}}_{ij}^{11}(0,\bq)-
\sum_{\alpha\gamma} \widetilde{\boldsymbol{\mG}}^{1\alpha}_{ix}(|x|,\bq)\widetilde{\boldsymbol{\mathcal{Q}}}^{-1}_{\alpha,\gamma}(\bq)\widetilde{\boldsymbol{\mG}}^{1\gamma}_{xj}(|x|,-\bq)\right],
\end{align}
where $\bq$ is the in-plane momentum and $\widetilde{\boldsymbol{\mG}}^{\alpha,\beta}$ is the inverse of $\widetilde{\boldsymbol{\mG}}^{-1}_{\alpha,\beta}$ given by Eq.~(\ref{eq:propagator}). $i,j$ stand for $x,y$ and $\boldsymbol{\mathcal{Q}}$ is given by Eq.~(\ref{eq:Q}). 
It can be shown that the displacement correlations on scales larger than $L_v$ are given by 
\begin{align}\label{eq:correlationrelevant}
\overline{\langle u_i(\br) u_j(\br)\rangle}&=\lim_{n\to 0}T\left[\mG_{ij}^{11}(0,\mathbf{0})-(\hat{\bx}\cdot\hat{\mathbf{i}})(\hat{\bx}\cdot\hat{\mathbf{j}})\mG_{xx}^{11}(2|x|,\mathbf{0})\right].
\end{align}
\end{widetext}
Using this result, we obtain for the
average change in the FL density 
\begin{align}
\overline{\langle\delta\rho(\br,\bu(\br))\rangle}=2\rho_0\sum_{m>0}\cos{(mG_Dx)} \left(\frac{L_v}{|x|}\right)^{m^2g} \, .
\end{align}
These oscillations resemble Friedel oscillations which can be also
found in Luttinger liquids with an isolated impurity\cite{Egger+95} or
in classical 2-dimensional systems with a columnar
defect\cite{Hofstetter+04,Affleck+04}.

The amplitude of the Friedel oscillations decays as a power law with
an exponent $g$ and $2g-1$ for a relevant and an irrelevant defect,
respectively. For an irrelevant defect the amplitude decays more rapidly than for a
relevant defect. In the absence of point impurities the defect is always
relevant in the RG sense, and the amplitude of the Friedel
oscillations remains finite for $|x|\to\infty$.

\section{Finite density of weak defects}
\label{section:weak}

\subsection{Model}
\label{sec:model_weak}

In this section we consider a finite density of parallel planar defects
with {\it random position}. We assume that defects extend along the
entire sample and are aligned parallel to the applied magnetic field.
There is a competition between the two random potentials from planar
defects and point impurities. The defects tend to localize the FLs and
hence favor order along the defect planes while point impurities
promote FL wandering.

The Hamiltonian reads
\begin{align}
\label{eq:Ham}
 {\cal H}=\cH_0+\int d^3r \big[V_{\rm P}(\textbf{r})+ V_{\rm
D}(\textbf{r})\big]\rho(\textbf{r},\textbf{u})\, ,
\end{align}
where $\cH_0$ is the elastic Hamiltonian of Eq.~(\ref{eq:elastic Ham})
and $V_P$ is the pinning potential resulting from point impurities,
see Eq.~(\ref{eq:pointpinning}). The defect pinning potential is
$
V_D(\textbf{r})= -v\left\{\sum_i\delta_\xi(x-x_i)-1/\ell_D\right\}
$
where we assumed that the
defect planes are parallel to the $yz$-plane and $\ell_D$ is a mean defect spacing. The $\delta$-function is
assumed to have a finite width of the order of the superconductor
coherence length $\xi$. The defect potential 
is uncorrelated along the $x$-axis,
\begin{align}
\overline{V_D(\br_1)V_D(\br_2)}=\frac{v^2}{\ell_D}\delta_{\xi}(x_1-x_2).
\end{align}
We discuss the case where the
gap between two defect planes typically contains many FLs, i.e.,
$\xi\ll\ell\ll\ell_D$. Note that orientation of the defects is
otherwise arbitrary.

After averaging over the defect positions, the replica Hamiltonian for the defects reads
\begin{widetext}\begin{align}\label{eq:replicaHD}
\cH^n_D=-\frac{(v\rho_0)^2}{2T\ell_D}&\int \dif^3\br_1\dif^3 \br_2 \sum_{\alpha,\beta} \delta_{\xi}(x_1-x_2)\Bigg\{-2\bnabla_{\bx}\bu^{\alpha}(\br_1)\sum_{\bG}e^{i\bG[\bx_2-\bu^{\beta}(\br_2)]}\notag \\&+\bnabla_{\bx}\bu^{\alpha}(\br_1) \bnabla_{\bx}\bu^{\beta}(\br_2) + \sum_{\bG_1,\bG_2}e^{i\bG_1[\bx_1-\bu^{\alpha}(\br_1)]}e^{i\bG_2[\bx_2-\bu^{\beta}(\br_2)]}\ \Bigg\}
\end{align}
where $\bx=(x,y)$. The defects are assumed to be sufficiently weak so that terms of the order $v^3/\ell_D$ and higher can be neglected.
The first term does not contribute to $\cH_D^n$ due to the oscillatory factor $e^{i \bG \bx_2}$ and the third term contributes only for reciprocal vectors perpendicular to the defects satisfying $\bG_1=-\bG_2=n \bG_D$ with integer $n$. Introducing the relative coordinate $x_r=x_1-x_2$ and taking into account that $\delta_{\xi}(x_r)$ is finite for $|x_r|\leq\xi$, we approximate the displacement field as $u_x(x_2+x_r,y_1,z_1)\approx u_x(x_2,y_1,z_1)$. This approximation is justified since the displacement field varies slowly over the FL spacing. Then Eq.~(\ref{eq:replicaHD}) can be written as
\begin{align}\label{eq:replicaHam}
 \cH^n_D=-\frac{1}{2T}\int_{x_1,y_1,z_1,y_2,z_2}\sum_{\alpha,\beta} \left\{  \sigma\bnabla_{\bx}\bu^{\alpha}(x_1,y_1,z_1) \bnabla_{\bx}\bu^{\beta}(x_1,y_2,z_2)   +R_D\left[u_x^{\alpha}(x_1,y_1,z_1)-u_x^{\beta}(x_1,y_2,z_2)\right]\right\},
\end{align}\end{widetext}
where
\begin{align}\label{eq:RD}
R_D(u_x)=\sigma\sum_{n\neq0} {\delta}_{\xi^{-1}}(n G_D)  e^{i n G_D u_x}
\end{align}
and we defined $\int_x=\int \dif x$ and
$\sigma=(v\rho_0)^2/\ell_D$. After averaging over point impurities,
the complete replica Hamiltonian is $\cH^{n}=\cH^{n}_{P}+\cH^n_D$,
where $\cH^{n}_P$ is given by Eq.~(\ref{eq:replica H}).

The first term in Eq.~(\ref{eq:replicaHam}) comes from the coupling of
the defect potential to the slowly varying part of the FL density
$\sim\bnabla_{\bx}\bu$. This term does not contribute to the glassy
properties of the system, since it can be eliminated by a simple
transformation\cite{Hwa+94},\footnote{Since the defects are weak, the
  effects we are interested in become visible on large length
  scales. Hence, we can introduce a coarse grained version of the
  defect potential $\widetilde{V}(x)=\int \dif x' V_D(x')/L_w$, where the integration is over a segment of length $L_w\gg \ell_D$. In that case the central limit theorem shows that
  $\widetilde{V}(x)$ is Gaussian distributed with
  $\overline{\widetilde{V}(x)\widetilde{V}(x')}=v^2/\ell_D \delta(x-x')$. In the
  defect Hamiltonian, the slowly varying part of the FL density
  couples only to $\mu(x)=\int_{|q_x|\sim 0}\widetilde{V}(q_x)\exp[i
  q_x x]/(2\pi)$ and to the periodic part
  $W(x)=\sum_{n}\int_{|q_x|\sim 0}\widetilde{V}(n G_D+q_x)\exp[i
  (q_x+n G_D) x]/(2\pi) $, where $n\neq 0$ is integer.  Since one has
  for the Gaussian potential $\overline{\widetilde{V}(q)\widetilde{V}(q')}=2\pi
  v^2/\ell_D \delta(q+q')$, the potentials $\mu(x)$ and $W(x)$ are not
  correlated, $\overline{W(x)\mu(x)}=0$. By applying the
  transformation $u'(\br)=u(\br)-\int_0^x \dif x_1 \mu(x_1)/c_{11}$
  and averaging over $W(x)$, we get only the second term of the
  replica Hamiltonian of Eq.~(\ref{eq:replicaHam}) and the first term
  has been eliminated.}. (For a more detailed discussion of this term
see below.) The remaining part of the replica pinning energy $\cH^{n}$
is invariant under the transformations
\begin{align}\label{eq:symmetry1}
u^{\alpha}_x(\br)\to u^{\alpha}_x(\br)+f_x(x),\\
u_y^{\alpha}(\br)\to u_y^{\alpha}(\br)+f_y(\br)\label{eq:symmetry2},
\end{align}
where $f_x(x)$ and $f_y(\br)$ are arbitrary functions.
Eq.~(\ref{eq:symmetry1}) represents an approximate symmetry if the
defect potential has a finite width. However, with increasing
  length scale, deviations from the symmetry become less important.
These symmetries show that the elastic coefficient $c_{11}$ is not
renormalized. Not renormalized are also the elastic moduli which
determine the energy cost for tilting the FLs \emph{only} in the $y$
direction (i.e.~the term $c_{44} (\partial_z u_y)^2$) and for changing
only the displacement $u_y$ along the $x$-axis (i.e.~the term $c_{66}
(\partial_x u_y)^2$).  These symmetries are commonly denoted as
statistical tilt symmetry\cite{Schulz+88}. However, the defects are an
important source of anisotropy and other elastic properties of the FLL
will be affected. For example, the energy cost for FL tilting as well
as FLL shearing parallel and perpendicular to the planes will differ
considerably. Also, due to the defect planes the system is not
invariant under arbitrary rigid rotations of the FLL around $z$ axis
and rotational modes will appear in the elastic Hamiltonian under
renormalization\cite{Kogan+89}. Note that for the FLL with point
disorder only, none of the elastic constants will be renormalized
since the disorder correlation function $R_P(\bu)$ is invariant under
the more general transformation $\bu^{\alpha}(\br)\to
\bu^{\alpha}(\br)+\mathbf{f}(\br)$.

Planar defects in the form of twin boundaries that are perpendicular
to the copper oxide planes very often appear in YBa$_2$Cu$_3$O$_{7-x}$
(YBCO)\cite{Sanfilippo+97,Kwok+92,Crabtree}. YBCO is a high
temperature superconductor and within high accuracy it is uniaxially
anisotropic\cite{Blatter+94}. YBCO can be reasonably well described
within a continuum anisotropic model, while for more strongly layered
superconductors a different description is needed. The elastic
description for anisotropic superconductors can be found in the review
article by \citet{Blatter+94}. The number of independent elastic
moduli increases with respect to the isotropic case that we discussed
in Sec.~\ref{section:BG}. However, if the magnetic field is applied
perpendicular to the copper oxide planes, the model given by
Eq.~(\ref{eq:elastic Ham}) as well as the considerations in the
following sections are directly applicable also for YBCO.

\subsection{Functional renormalization group approach}
\label{section:FRG}

In the previous section we treated a single defect plane as a
perturbation to the Bragg glass fixed point. Now, we consider both the
planar defects and the point impurities as a perturbation to the ideal
Shubnikov phase. Notice that Eq.~(\ref{eq:RD}) depends only the
displacement field $u_x$. Since we focus below on the effect of the
defect planes, it seems to be justified to start from a simplified
model in which only the displacement $u_x\equiv u$ of the FLs \emph
{perpendicular} to the defect planes is considered. This model
describes also a wide class of other systems which exhibit regular
lattices of domain walls like magnets, charge density waves
\cite{gruener} and incommensurate systems \cite{BrCo78}.

In the absence of the defect planes point impurities are relevant
below four dimensions. We employ here an Imry--Ma--type argument
\cite{ImMa} in combination with perturbation theory to see the effect
of randomly distributed point impurities on the FLL.  When the initially
ordered FLL is distorted in a volume $L^d$ by $u\sim\ell$, the typical
energy gain is of the order $\sim(-R_P''(0)L^d)^{1/2}$ compared to the
elastic energy loss $\sim L^{d-2}$. For $d<d_P=4$ and sufficiently
large $L\gg L_{\xi}\sim \left(-R_P''(0)\right)^{1/(d-4)}\gg \ell$ the
point disorder wins and the FLL becomes distorted.  A more detailed
study shows that in this case the FLL exhibits a phase with quasi long
range order which is the previously discussed Bragg glass phase (see
Sec. \ref{section:BG}). In this phase the positional correlation
function $S_{\mathbf{G}}$ [see Eq.~(\ref{eq:positional correlation})]
shows a power law decay.

Next we consider the Imry-Ma argument for planar defects in a volume
$L_x^{d-2}L_zL_y$ without point impurities.  The energy gain is of the
order $ (-R_D''(0) \ell^2 L_x^{d-2})^{1/2}L_zL_y$. The elastic energy
loss is $ c_{11}L_zL_yL_x^{d-4}\ell^2 $ since distortions are aligned
parallel to the defects. For $ L_x\gg L_D\sim
(-c_{11}^2\ell^2/R_D''(0))^{1/(6-d)} $ the pinning energy gain wins
and the FLL becomes disordered in the direction perpendicular to the
defects. $L_D$ is the so-called Larkin length for the defects.  The
critical dimension above which weak planar defects are irrelevant is
$d_D=6$.

For an RG approach is convenient to consider a generalization of our
model to $d$ dimensions. The defects remain two-dimensional with $d-2$
transverse directions, while the displacement field remains
uniaxial. In the following, we use a functional RG
approach\cite{Fisher86,Balents+93} in $d=6-\epsilon$ dimensions. We
follow closely a related approach for columnar disorder
\cite{Balents,Fedorenko} but do not rescale the renormalized
quantities so that they correspond to the effective parameters
measured on the scale $L_x$. Thermal fluctuations and point disorder
are irrelevant for $\epsilon<4$ and $\epsilon<2$, respectively. Hence
we can assume directly $T=0$ and $R_{P}=0$. To lowest
order in $\epsilon$ the RG flow equation read
\begin{align}\label{eq:RG1}
&\frac{\dif\ln c_{ii}}{\dif\ln L_x}=
\frac{K_dR_D''''(0)L_x^{\epsilon}}{c_{11}^2},
\quad i=4,6\\\label{eq:RG2}
&\frac{\dif R_D(u)}{\dif \ln L_x} =
\frac{K_dR_D''(u)L_x^{\epsilon}}{2c_{11}^2}\big[R_D''(u)-2R_D''(0)\big],
\end{align}
where $K_d={S_{d-2}}/{(2\pi)^{d-2}}$ and $S_{d}$ denotes the surface
of the $d$-dimensional unit sphere.

In a static situation the displacement field is independent on $y$ and
$z$ since defects distort FLL planes that are parallel to the
$yz$-plane on the whole. Since, by assumption, other sources of
fluctuations are not present we can perform the integration over $y$
and $z$ in the Hamiltonian of Eq.~(\ref{eq:replicaHam}) and
obtain an effective $d-2$ dimensional Hamiltonian that describes the
interaction of FLL planes with defect planes with $d-2$ dimensional random positions. This explains why the flow equation
for $R_D$ has the form as the one for the point disorder correlator
$R_P$ in $d-2$ dimensions\cite{Giamarchi+94}. However, an important
difference between the $d$ dimensional FLL with defects and the FLL
planes with point impurities in $d-2$ dimensions is the
renormalization of elastic constants $c_{44}$ and $c_{66}$ in the
former model.

For $L_x\to L_D$, $R_D^{''''}(0)$ increases and at $L_x\approx L_D$
$R_D^{''}(u)$ develops a cusp at $u=0$. The cusp signals the appearance of 
metastable states the energy of which is very close to the ground
state energy but which may be far apart in configuration space. The cusp
results in diverging elastic constants $c_{44}$ and $c_{66}$ and in a
change of the sign of $R''''(0^+)$ from positive to negative. If there
is a small but finite tilt or shear of FLs, $R_D^{''''}(0)$ has to be
replaced by $R''''(0^+)$ in Eq.~(\ref{eq:RG1}), and on length scales
$L_x>L_D$ the elastic constants decrease since $R''''(0^+)$ is then
negative. Importantly, a new term of the form
\begin{align}\label{eq:newterm}
 \frac{1}{\ell}\int d^{d-2}x
 dy dz\left|{\Sigma_y}(\partial_{y}u)\hat{y}+
 {\Sigma_z}(\partial_{z}u)\hat{z}\right|
\end{align}
is generated in the Hamiltonian. $\Sigma_{z(y)}$ has the meaning of a
interface tension of a domain wall perpendicular to z (y) axis, across which the displacement field changes by $\ell$. They dominate the elastic energy
for small $u$ and are renormalized according to
\begin{align}
  c_{66}^{-1/2}\frac{\dif \Sigma_y}{\dif \ln L_x}=
  c_{44}^{-1/2}\frac{\dif \Sigma_y}{\dif \ln L_x}=R_D^{'''}(0^+)
  L_x^{5-d}\frac{K_d\ell}{2c_{11}^{3/2}} \, .
\end{align}
$\Sigma_z$ and $\Sigma_y$ satisfy the relation
$\Sigma_z/\Sigma_y=\sqrt{c_{44}}/\sqrt{c_{66}}$. Notice that $c_{44}$
and $c_{66}$ are renormalized in the same way such that their ratio
remains constant under the RG flow.

Next we estimate the Larkin length $L_D$ from the flow equations. The
function $R_D(u)$ is even and as long as it is analytic, all odd
derivatives at $u=0$ vanish. Assuming analyticity, the flow equation
for $R_D^{''''}(0)$ reads
\begin{align}
\frac{d R_D^{''''}(0)}{d \ln L_x}=\frac{4}{c_{11}^2}K_d L_x^{\epsilon} [R_D^{''''}(0)]^2\, .
\end{align}
The solution on the length scale $L_x$ is given by
$R_D^{''''}(0,L_x)=R_D^{''''}(0,\lambda)\left[1-R_D^{''''}(0,\lambda) 4K_d \left( L_x^{\epsilon}-\lambda^{\epsilon} \right) /c_{11}^2 \epsilon \right]^{-1}$,
where $\lambda$ is the penetration depth and has a role of the small length scale cutoff. This shows that $R_{D}(0)$ diverges at
\begin{align}\label{eq:L_D}
L_D\approx \left[\frac{\epsilon c_{11}^2}{4K_d R_{D}^{''''}(0,\lambda)}\right]^{1/\epsilon} \, .
\end{align}
This result is in qualitative agreement with the estimate we obtained
from scaling arguments since $R_D^{''''}(0,\lambda)\ell^2\sim
R_D^{''}(0,\lambda)$.

The fixed point function has for $0\le u<\ell$  the
form\cite{Giamarchi+94}
\begin{align}\label{eq:fixed}
R_D^{*''}(u,L_x)=-\frac{\epsilon
    c_{11}^2L_x^{-\epsilon}} {6K_d}\left[\left(u-\frac{\ell}{2}\right)^2-\frac{\ell^2}{12}\right]
\end{align}
and it has to be periodically continued in $u$ with period
$\ell$. Note that if we would consider rescaled quantities then in
Eq.~\eqref{eq:fixed} $L_x^{-\epsilon}$ would be replaced by
$\Lambda^{\epsilon}$. In the case of a finite tilt (shear) for $L_x>L_D$
the elastic constant $c_{44}$ ($c_{66}$) decreases as $c_{ii}(L_x)\sim
({L_x}/{L_D})^{-\epsilon/3}$. A rough estimate for the saturation
value for the interface tension is $\Sigma_z\sim\epsilon
\ell^2\sqrt{c_{11}c_{44}(\lambda)}/L_D$.

As has been pointed out by Fedorenko\cite{Fedorenko}, the RG equations
for the elastic constants $c_{44}$, $c_{66}$ and the interface
tensions resemble those of the friction and driving force
for the depinning transition of the FLL in the presence of point
impurities\cite{Nattermann+92}. The role of velocity is here played by
tilt or shear and the elastic constants diverge at zero tilt and zero
shear as the friction diverges in the static case. Also, as a
threshold force exists for the depinning transition, the interface tension
$\Sigma_{z(y)}$ determines the threshold force for tilting (shearing)
the FLL, as we will see in the next subsection.
\subsection{Properties of the planar glass}
\label{section:properties}

In this subsection we summarize the properties of the new phase that
is described by the fixed point of the functional RG of the previous
section and in the following is called ''planar glass''.  We examine
the response of the system to FL tilting, to a change in the
longitudinal magnetic field and discuss the order of the FLL. We show that
the new phase and its properties are robust against weak point
  impurities in $d=3$ and $d=4$.
 
When one changes the direction of the applied magnetic field by $ H_x
\mathbf{ \hat x}$, the Hamiltonian changes by
\begin{align}
 \delta \cH=-\frac{\phi_0\rho_0}{4\pi}
\int d^3r\;
 H_x\partial_{z}u .
\end{align}
To tilt the FLs with respect to the $z$-axis, $H_x$ has to overcome
the interface energy $\sim\Sigma_z$ which results in a threshold field
\begin{align}\label{eq:treshold field}
H_{x,c}=2\pi\sqrt{3}\frac{\Sigma_z a^2}{\phi_0\ell}
\end{align}
below which the FLs remain locked parallel to the planes. This is the
\emph{transverse Meissner effect}: a weak transverse magnetic field
$H_x$ is screened from the sample and $c_{44}$ is infinite.
Only for $ H_x> H_{x,c}$ the average tilt of the FLs becomes non-zero
and $c_{44}$ is finite. In this way, by measuring the threshold field,
$\Sigma_z$ can be measured.

Moreover, there is a \emph{resistance against shear} of the FLL. The
shear deformation $\partial_yu_x$ is non zero (and $c_{66}$ is finite)
only if the shear stress $\sigma_{xy}$ is larger than a critical value
$\sigma_{xy,c}=\Sigma_y/\ell$.  Otherwise $c_{66}$ is infinite.  The
divergence of $c_{66}$ is a new property that does not appear in the
Bose glass which, however, does also show a transverse Meissner effect.

An infinitesimal change in the longitudinal magnetic field $\delta
H_z\mathbf{\hat z}$ changes the Hamiltonian by
\begin{align}
 \delta \cH=-\frac{\phi_0\rho_0}{4\pi}
\int d^3r\;
 \delta H_z\partial_{x}u
\end{align}
and allows to measure the longitudinal magnetic  susceptibility
$\chi=\phi_0\rho_0\partial\langle\partial_x u \rangle/\partial \delta
H_z$.  The disorder averaged susceptibility is
\begin{align}
 \overline{\chi}=\frac{(\phi_0\rho_0)^2}{4\pi c_{11}},
\end{align}
as shown in Appendix \ref{app2}. It is independent of disorder as
a result of the statistical tilt symmetry\cite{Schulz+88}. The glassy
properties of the system can most easily be seen by the sample to
sample fluctuations of the magnetic susceptibility.
Perturbation theory yields (see Appendix \ref{app2})
\begin{align}
\frac{\overline{\chi^2}-{\overline
\chi}^2}{\overline
\chi^2}=\frac{R_D''''(0)L_x^{\epsilon}}{5c_{11}^2}\sim
\left(\frac{L_x}{L_D}\right)^{\epsilon} \, ,
 \end{align}
 i.e.~the sample to sample fluctuations of the susceptibility grow
 with the scale $L_x\lesssim L_D$, $d<6$. We cannot expect that this
 result is quantitatively correct for large $L_x$, but qualitatively
 it demonstrates the relevance of defects and it provides a signature
 of a glassy phase\cite{Hwa+94}. Although we were not able to prove
 it, $\overline{\chi^2}/{\overline \chi}^2-1$ will most likely
 approach a finite universal value for $L_x\gg L_D$ in $d<6$.

 The positional correlation function is obtained to first order in
 perturbation theory, combined with a functional RG analysis for
 $6>d>4$. It reads
\begin{align}\label{eq:logarithmic_correlation}
S_{\bf G_D}({\bf x},y,z) \sim |{\bf x}|^{-\eta_D}
\end{align}
where $\eta_D=(\pi/3)^2(6-d)$.  A detailed derivation of this result
is presented in Appendix \ref{app3}. In order to study the behavior in
$d\leq 4$, we have to reconsider the first term of
Eq.~(\ref{eq:replicaHam}). It results from the coupling of the defect
potential to the slowly varying part of the FL density
($\sim \partial_x u$).  By taking into account that at a scale $L_x$
the displacement field behaves as $u\sim L_x^{\zeta}$, we find that
the $\sigma$-term scales as $\sim L_y^2L_z^2 L_x^{d-4+2\zeta}$. When
we compare the latter term to the squared elastic Hamiltonian $\sim
L_y^2L_z^2 L_x^{2(d-4+2 \zeta)}$ that describes the cost of deviations
of $u$ in the $\bx$ directions only (since the FLs are completely
ordered parallel to the defects on sufficiently large scales in the
absence of point impurities), we find that the $\sigma$-term becomes
relevant if $d-4+2\zeta \leq 0$. For logarithmic roughness ($\zeta=0$)
it is relevant for $d\leq 4$. Since the other part of the defect
pinning energy $\sim R_{D}(u)$ scales in the same way as the elastic
energy, the $\sigma$-term is the dominant part of the pinning energy
and determines the FL roughness.

First, we consider the case without point impurities and then treat
them perturbatively. Applying a Flory--type argument \cite{Natter+00},
i.e., assuming that the elastic energy and the dominant part of the
defect energy scale in the same way, we find, following the discussion
above, that in $d=3$ the roughness exponent is $\zeta=1/2$. More
detailed calculations \cite{ViFE84,Fe80} confirm our result, leading
to
\begin{align}
S_{\bf G_D}({x},y,z) \sim e^{-|x|/\xi_c} \, ,
\end{align}
where $\xi_c\approx L_D$. Note that there is a shift of dimension
$d\to d+2$ between the model studied in Ref.~\cite{ViFE84,Fe80} to our
model since the FLs are ordered in the $yz$-plane. In
Ref.~\cite{ViFE84,Fe80} a related one-dimensional system with point
impurities at zero temperature is studied. There is a nontrivial
renormalization of $\sigma$ coming from the defect potential that
couples to the periodic part of the FL density\cite{ViFE84}.  The
$\sigma$-term does not contribute to the renormalization of $R_D$,
since the $\sigma$-term can be eliminated in every step of the RG
procedure by applying the transformation that
does not affect the correlator $R_D$, as discussed at the beginning of
this section. That is why $\Sigma_z$ and $\Sigma_y$ will be generated
also for $d\leq4$. Villain and Fernandez\cite{ViFE84} found from a
non-perturbative RG that for $d\le 4$ the defect-induced disorder
flows under the RG to strong coupling. However, our study of the
strong coupling limit in Section IV shows that this limit gives
qualitatively the same result as the case investigated in this
section.

To summarize, the planar glass phase is characterized by (i) diverging
shear and tilt moduli but a finite compressibility, (ii) a transverse
Meissner effect as well as a resistance against shear deformation,
(iii) sample to sample fluctuations of the longitudinal magnetic
susceptibility and (iv) an exponential decay of positional
correlations in the direction perpendicular to the defects in $d=3$.

Since point disorder may formally become relevant below $d=4$, we
consider the stability of the planar glass with respect to weak
point impurities. We find that the pinning energy due to point
impurities in $d=3$ behaves as
\begin{align}\label{eq:1}
\overline{\langle \cH_P \rangle}&=\rho_0\sum_{n\neq0}\int \dif \br V_P(\br) e^{i n G_D x}e^{-(n G_D)^2\overline{\langle u^2\rangle}/2}\notag\\&\sim \rho_0 \sqrt{n_{imp}v_p^2 L_yL_zL_{x}}  e^{-L_x/(2\xi_c)}
\end{align}
and from this we conclude that weak point impurities are an irrelevant
 perturbation. Similarly, it can be shown that pinning energy of randomly
  distributed {\it columnar} defects decays exponentially with $L_x$ and hence does
not destroy the planar glass.

\subsection{Stability of the Bragg glass and the weakly pinned Bose glass}

In this subsection we continue discussion of the competition between
pinning effects due to point impurities and columnar and planar
defects. We shall show that the weakly pinned Bose glass is stable
with respect to weak point impurities but unstable with respect to
weak planar defects. Moreover, we shall demonstrate that the Bragg
glass phase is unstable with respect to both weak planar and weak
columnar defects. The resulting phase diagram is shown schematically
in Fig.~\ref{Fig:1}.

First we discuss the stability of the Bragg glass phase in analogy to
the test for stability of the planar glass in the previous
subsection. We note that the correlation functions in the Bragg glass
phase can also be obtained from a model with a uniaxial displacement
field of FLs\cite{Giamarchi+95}. A uniaxial displacement field
describes also properly charge density waves, a stack of membranes
under tension and domain walls in magnets. Justified by these
observations, we first consider a uniaxial displacement field in the
direction perpendicular to the defect planes. At the Bragg glass fixed point in $d=3$ the pinning energy of the planar defects
behaves as
\begin{align}\label{eq:2}
\overline{\langle \cH_D \rangle}&=\rho_0\sum_{n\neq0}\int \dif \br V_D(\br) e^{i n G_D x}e^{-(n G_D)^2\overline{\langle u^2\rangle}/2}\notag\\&\sim \rho_0\sqrt{v^2 L^5/\ell_D}L^{-\pi^2/18}.
\end{align}
When we compare this energy to the pining energy of point impurities,
\begin{align}\label{eq:bgscaling}
\overline{\langle \cH_P \rangle} & \sim \sqrt{L^3 R_P^*(0)}\sim L \, ,
\end{align}
we find that planar defects are a relevant perturbation. Note that at
the Bragg glass fixed point the system is isotropic, i.e.,
$L_x=L_y=L_z=L$. In Eq.~(\ref{eq:bgscaling}) we used the fact that the
fixed point correlator on the length scale $L$ behaves as
$R_P^*(0)\sim L^{-1}$ [cf.~Eq.~(\ref{eq:correlator})]. Similarly, the
pinning energy of columnar defects at the Bragg glass fixed point
scales as
\begin{align}
\overline{\langle \cH_C \rangle}
&\sim \rho_0\sqrt{v_c^2n_{cd} L^4}L^{-\pi^2/18},
\end{align}
and drives the system away from the Bragg glass fixed point.

A functional RG analysis of weak columnar defects in $d=5-\epsilon$
yields a stable phase with a zero temperature fixed point that is
characterized by a power law decay of the positional correlation
function with an exponent $\eta_C=(\pi/3)^2(5-d)$ and a transverse
Meissner effect\cite{Balents93}. One can expect that this phase, found
for an uniaxial displacement field, applies to the case where the FL density is larger than the columnar defect density, corresponding to the so called weakly pinned Bose glass. 
In order to 
study the stability of this phase in $d=3$ with respect to planar
defects and point impurities, we compare the scaling
\begin{align}
\overline{\langle \cH_D \rangle}&\sim \rho_0\sqrt{v^2 L^3L_z^2/\ell_D}L^{-(\pi/3)^2},\\
\overline{\langle \cH_C \rangle}&\sim \sqrt{R_c^*(0)L^2L_z^2}\sim L_z,\\
\overline{\langle \cH_P \rangle}&\sim \rho_0\sqrt{v_p^2 n_{imp} L^2L_z}L^{-(\pi/3)^2}
\end{align}
of the different pinning energies, where we used $L_x=L_y=L$ and
columnar disorder fixed point correlator $R_C^*(0)\sim L^{-2}$. We conclude that weak point impurities are
irrelevant but weak planar defects are relevant in the weakly
  pinned Bose glass.

Next we examine the stability of the Bragg glass phase by considering
a vector displacement field. This displacement field reveals the triangular lattice structure of
 the FLL. By changing the orientation of the defect planes, the number of FLs that
are pinned by the defects changes. In $d=3$ we have
\begin{align}
\overline{\langle \cH_D \rangle}&\sim \rho_0\sqrt{v^2 L^5/\ell_D}e^{-{G_D^2}\overline{\langle u_x^2 \rangle}/2 }\sim L^{\frac{5}{2}-g}, \\
\overline{\langle \cH_C \rangle}&\sim \rho_0\sqrt{v_c^2 n_{cd} L^4} e^{-\overline{\langle (\mathbf{G}_0 \mathbf{u})^2 \rangle}/2 }\sim L^{2-\eta/2}, \\
\overline{\langle \cH_P \rangle} & \sim \sqrt{L^3R_P^*(0)}\sim L,
\end{align}
where $g$ is given by Eq.~(\ref{eq:flow}),
$|\mathbf{G}_0|=4\pi/(\sqrt{3}a)$ is the shortest vector of the
reciprocal lattice and $\eta$ is the exponent of the positional
correlation function in the Bragg glass phase. Weak columnar defects
are always a relevant perturbation, while weak planar defects are
relevant only if they satisfy $g<g_c=3/2$, i.e., only if they are
parallel to the main crystallographic planes of the FLL. Here we
neglected the influence of weak planar defects on the elasticity of
the FLL. In fact, the defects lead to an additional anisotropy in the elastic energy which is associated with a larger energy for deformations with nonzero $\partial_y u_x$ and
$\partial_zu_x$. Through a renormalization of the elastic
constants $g$ is renormalized downwards. Therefore, stronger planar
defects lead to an increased $g_c>3/2$, rendering additional
orientations of defects relevant. However, it is likely that in the
case of a finite density of parallel defect planes, the FLL will
rotate to a position in which it reaches maximum overlap with the
defects. Then the planar defects will be parallel to the main lattice
planes and the Bragg glass is unstable.

\section{Finite density of strong defects}
\label{section:strong}

In this section we consider the FLL with a finite density of parallel,
randomly distributed defect planes that are aligned to the magnetic
field.  The defects are assumed to be sufficiently strong so that the
Larkin length [see Eq.~(\ref{eq:L_D})] is of the order of the mean
defect spacing or smaller, $L_D\lesssim \ell_D$. In this case
the defect potential can not be treated perturbatively with respect to
the elastic energy, and a new approach is required. Here we derive an
effective Hamiltonian that is defined only at the defect planes. We
determine the ground state configuration of the FLL and calculate the
positional correlation function. We show that a transverse Meissner
effect as well as a resistance against shear deformations appear also
in this case.

The Hamiltonian in $d=3$ dimensions reads
\begin{align}
\cH=\cH_0+\sum_{i}^{N_D}\cH_{D,i},
\end{align}
where $\cH_0$ is the elastic Hamiltonian given by Eq.~(\ref{eq:elastic
  Ham}) and $\cH_{D,i}$ is the pinning energy of the defect plane at
the position $x=x_i$, see Eq.~(\ref{eq:defect H}). $N_D$ denotes the
number of defects. In this section we consider a simplified model
involving only uniaxial displacements perpendicular to the defects
$\bu=u\hat{x}$. In Sec.~\ref{section:conclusions} we shall discuss the
implications of the generalization to a two-dimensional vector
displacement. The part of $\cH_D$ that describes the coupling of the
pinning potential to the slowly varying part of the FL density
($\sim \partial_x u$) leads  only to an increase in the FL
density at the defects and can be eliminated by applying the
transformation
\begin{align}\label{eq:transformation}
u(\br)\rightarrow u(\br)-{v\rho_0}/{c_{11}}\sum_{i=1}^{N_D}\Theta(x-x_i) \, ,
\end{align}
where $\Theta(x)=0,\;x\leq0$ and $\Theta(x)=1,\;x>0$.  The Hamiltonian
then becomes
\begin{align}\label{eq:hamiltonianstrong}
\cH=\cH_0-2v\rho_0\sum_{i=1}^{N_D}\int_{y,z}
\sum_{k>0}^{[{\ell}/{\xi}]_G}
\cos{\left\{kG_{D}\left[u(x,y,z)-\alpha_i\right]\right\}},
\end{align}
where $\alpha_i=x_i+{v \rho_0(i-1)}/{c_{11}}$. For simplicity we
assume that all defects have the same strength.

In order to obtain a Hamiltonian that is isotropic in the $yz$-plane,
we introduce the rescaled coordinate $z'=z\sqrt{{c_{66}}/{c_{44}}}$
and define $u'(y,z')=u(y,z)$. We shall omit the primes below. We
proceed by studying the ground state of the displacement field for a
given distribution of planar defects, assuming that a strong defect
potential suppresses thermal fluctuations. First we solve the saddle
point equation in the gap between the defects with prescribed, but
arbitrary, boundary conditions at the defects $u_i(y,z)=u(x_i,y,z)$
with Fourier transform $\widetilde{u}_i(\bq)=\widetilde{u}(x_i,\bq)$,
$\bq=(q_y,q_{z})$. In the ground state configuration the FLs are
completely aligned to the defect planes and $u(x_i,y,z)$ is
independent of $y$, $z$. However, we derive the saddle point solution
and the effective Hamiltonian for a more general displacement field
configuration at the defect planes since this will be necessary for a
discussion of the transverse Meissner effect below as well as the FL
dynamics in Sec.~\ref{sec:Creep}.  In the following we use the
notation $\Delta A_i=A_{i+1}-A_i$ for any quantity $A$. The solution
of the saddle point equation between two defect planes reads, with
$x\in[x_i,x_{i+1}]$, 
\begin{align}
\widetilde{u}(x,\bq)=\frac{\widetilde{u}_i(\bq)}{\sinh{(q' \Delta x_i)}}
\sinh{\left[q'(x_{i+1}-x)\right]}\notag\\+\frac{\widetilde{u}_{i+1}(\bq)}
{\sinh{\left(q' \Delta x_i\right)}}\sinh{\left[q'(x-x_{i})\right]},
\end{align}
where $q'=\sqrt{c_{66}/c_{11}}q$. Note that we have solved the saddle
point equation beetween the defects within a continuum model and not
on the lattice. This amounts to setting the momentum cutoff
$\Lambda\to \infty$.  After substituting this solution into the
Hamiltonian of Eq.~(\ref{eq:hamiltonianstrong}) and integrating over
$x$, the Hamiltonian reduces to
\begin{align}\label{eq:reducedH}
\cH=&\frac{\sqrt{c_{11}c_{44}}}{2}\int\frac{\dif^2q}{(2 \pi)^2}q
\sum_{i=1}^{N_D-1} \Bigg\{\frac{|\widetilde{u}_{i+1}(\bq)-\widetilde{u}_i(\bq)|^2}{\sinh\left(q
\Delta x_i\sqrt{\frac{c_{66}}{c_{11}}}\right)}\notag\\ & +\left(|\widetilde{u}_i(\bq)|^2+|\widetilde{u}_{i+1}(\bq)|^2\right)
\tanh{\left(\frac{q\Delta x_i}{2}\sqrt{\frac{c_{66}}{c_{11}}}\right)}
\Bigg\}\notag\\&-2v\rho_0\sqrt{\frac{c_{44}}{c_{66}}}\sum_{i=1}^{N_D}\int_{y,z}
\sum_{k>0}^{[{\ell}/{\xi}]_G}
\cos{\left\{kG_{D}\left[u_i(y,z)-\alpha_i\right]\right\}}.
\end{align}
A similar Hamiltonian has been obtained for a Luttinger liquid with
point impurities\cite{Malinin+07}.

Next we study the ground state of the FLL. In the limit $v\to\infty$
the FLs are completely aligned to the defects and $u_i(y,z)=u_i$.
Then the Hamiltonian of Eq.~(\ref{eq:reducedH}) becomes
\begin{align}
\label{eq:ground_hamiltonian}
\frac{\cH}{L^2}\sqrt{\frac{c_{66}}{c_{44}}}=&\frac{c_{11}}{2}\sum_{i=1}^{N_D-1} \frac{(u_{i+1}-u_i)^2}{\Delta x_{i}} \notag\\&-2v\rho_0\sum_{i=1}^{N_D}\sum_{k>0}^{[{\ell}/{\xi}]_G}
\cos{\left[k\frac{2\pi}{\ell}\left(u_i-\alpha_i\right)\right]}\, ,
\end{align}
where $L$ is the system size. For $u_i=\ell n_i+\alpha_i$, where $n_i$
is an integer number, the energy gain of the FLs from the defect
potential is maximal. We determine $n_i$ such that the elastic energy
is minimal and find the ground state configuration to be degenerate
and given by\cite{GlNa04}
\begin{align}\label{eq:ground}
u_i^{(n)}=\ell\left(\frac{\alpha_i}{\ell}-
\sum_{j<i} \left[\frac{\Delta\alpha_j}{\ell}\right]_G+n\right)\, ,
\end{align}
where $n$ is an integer number. Note that this is the ground state
configuration for an arbitrary defect strength in the special case
when ${\Delta \alpha_i}/{\ell}-
\left[{\Delta\alpha_i}/{\ell}\right]_G=0$ for all $i$. Then the FLs
are just shifted in order to gain energy from the defects without any
elastic energy loss.  However, for randomly distributed defect planes
that satisfy $\ell_D\gg\ell$,
${\Delta\alpha_i}/{\ell}-\left[{\Delta\alpha_i}/{\ell}\right]_G$ is
uniformly distributed in the interval $[-1/2,1/2]$. Using the central
limit theorem, we find that the positional correlation function decays
exponentially fast in the $x$ direction,
\begin{align}
S_{\bf G_D}({\br})\sim e^{-|x|/\xi_c},\quad\quad x\gg\ell_D
\end{align}
with $\xi_c\approx 6 \ell_D/\pi^2$. This shows that the limits of
weak and strong planar defects lead to the same behavior of the
positional correlations in $d=3$. The correlation length in both
cases is determined by the Larkin length of the defects.

From the shifted boundary conditions $u_i(y,z\to \infty)=u_i^0+\ell$
and $u_i(y,0)=u_i^{0}$ one can obtain also the interface tension
$\Sigma_z$ and it turns out to be finite. We do not quote the result
here since it is cutoff dependent and hence non-universal. Also,
a general
  expression that is valid for all ratios of the elastic constants is
  not available. A similar analysis shows that the surface tension
$\Sigma_y$ is finite. Hence, strong defects lead to a transverse
Meissner effect as well as a resistance against shear
deformations. The finite values of $\Sigma_{z}$ and $\Sigma_y$ 
might be interesting to probe experimentally.




\section{Flux Line Creep}
\label{sec:Creep}

The technologically the most interesting property of type-II
superconductors is their ability to carry a bulk current with as
little dissipation as possible. The Lorentz force acts on FLs and
hence gives rise to dissipation \cite{Tinkham}. Pining centers play an
important role in preventing FL motion and lead to a nonlinear
resistivity $\ln \rho\sim -J^{-\mu}$ for $J\to 0$ which depends on the
so-called creep exponent $\mu$ \cite{Blatter+94,Natter+00}. In this
section we study the effect of planar defects on the FL dynamics in
the direction perpendicular to the defect planes. The defects are
assumed to be parallel to the applied magnetic field. For a single
defect plane we show that the creep exponent is $\mu=1$, apart from
logarithmic corrections. We find that many planar defects act as a
more effective source of pinning than point
impurities \cite{Nattermann90} and columnar
defects \cite{NeVi+92,NeVi+93,Blatter+94}. They considerably slow down the FLs in
comparison to the Bragg glass and the Bose glass, leading to a creep
exponent $\mu=3/2$ for the planar glass.

\subsection{Single defect}\label{section:creep_single}
\begin{figure}
 \includegraphics[width=0.4\linewidth]{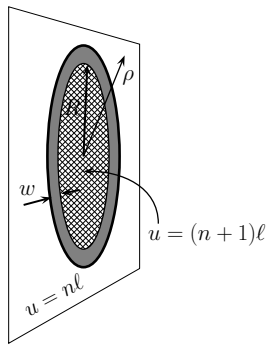}
 \caption{Schematic illustration of a droplet with radius $R$ and
   width $w$ at a defect plane.}\label{Fig:single}
 \end{figure}
 First we discuss the creep of FLs in the presence of a single
 planar defect, aligned to the applied magnetic field and without
 point impurities, for currents parallel to the defect and
 perpendicular to the magnetic field. The Hamiltonian is given by
 $\cH=\cH_0+\cH_D+\cH_{force}$, where the part describing elastic
 deformations, $\cH_0$, is given by Eq.~(\ref{eq:elastic Ham}), the
 defect pinning energy $\cH_D$ by Eq.~(\ref{eq:defect H}) and
 the Lorentz force contribution reads
\begin{align}\label{eq:lorentz_hamiltonian}
\cH_{force}=-\int \dif^3\br \left\{\mathbf{J}(\br)\times\mathbf{B}(\br)\right\}\cdot\bu(\br)\, .
\end{align}
$\mathbf{J}(\br)$ is the current density, $\mathbf B(\br)$ is the
magnetic induction and the speed of light is set to one. Since we will consider
only FL dynamics normal to the defect and since the defect potential
depends only on the perpendicular displacement field, it appears
plausible to simplify our model and to consider a uniaxial
displacement field in the direction perpendicular to the defect plane.

The defect plane is a relevant perturbation in the Shubnikov (mixed)
phase for all orientations of the plane (in contrast to the case when
point impurities are present) since thermal fluctuations do not
roughen the FLL in $d=3$. This can be easily seen from the RG equation
(\ref{eq:RG-defect-flow}) by setting $g=0$.  Under renormalization the
weak defect potential that couples to the periodic part of the FL
density grows and flows to strong coupling. Since we are interested in
the small current densities $J\to 0$ which probe large length scales,
we shall study only the strong defect plane below. Our results apply
also to weak defects at sufficiently large length scales or small
currents.

In the absence of a current, the FLs are aligned to the defect plane
and the ground state is highly degenerate. Different ground states
differ by a shift of the displacement field by $\ell$ because the
energy does not depend on which FLL plane is pinned by the
defect. This degeneracy is broken when a current is turned on and the
original ground state becomes unstable. The system now evolves into a
new metastable state that is lower in energy and in which the FLs are
shifted by $\ell$. This process is enabled via the formation of
droplets which are nuclei of new metastable states. The competition between bulk energy
gain and elastic energy loss determines the energy (and size) of the
critical droplet. The energy of the
critical droplet corresponds to the energy barrier that the FLs have
to overcome when evolving to a new state. Thermally activated FL
hopping over barriers with energy $E_{drop}^*(J)$ determines the
resistivity through the Arrhenius law {\cite{Van'tHoff,Arrhenius}}
\begin{align}
\rho(J)\sim e^{-{E_{drop}^*(J)}/{T}}.
\end{align}
Therefore, we need to estimate $E_{drop}^*(J)$.

We proceed by deriving an effective Hamiltonian that is defined on the
defect plane ($x=0$). By integrating out the displacement field off
the defect we get (see Appendix \ref{app:effective})
\begin{align}\label{eq:effectiveH}
\cH_{eff}&=\frac{1}{2}\int\frac{\dif^2 \bq}
{(2 \pi)^2}|\widetilde{u}(0,q_y,q_z)|^2 [\widetilde{\mathcal{G}}(0,\bq)]^{-1}
\notag\\&- 2v\rho_0\int \dif
y\dif z\sum_{k>0}^{[{\ell}/{\xi}]_G}
\cos{\big[kG_{D}u(0,y,z)\big]}\notag\\
&-\widetilde{u}(0,\bq=0)\int \dif x {J(x)B(x)}\, .
\end{align}
Here $\bq=(q_y,q_z)$ is the in-plane momentum, $\widetilde{u}(0,q_y,q_z)=\widetilde{u}(x=0,q_y,q_z)$ and
\begin{align}\label{eq:effectivepropagator}
\widetilde{\mathcal{G}}(0,\bq')=\frac{\arctan\left[\frac{\Lambda}{|\bq'|}\sqrt{\frac{c_{11}}{c_{66}}}\right]}{\pi |\bq'|\sqrt{c_{11}c_{66}}}\, ,
\end{align}
where $\bq'=(q_y, \sqrt{c_{44}/c_{66}}q_z)$ and
$\Lambda=2\pi/\lambda$. Since the system is translationally invariant
in the $yz$-plane, the current density and magnetic induction depend
only on $x$.  In order to simplify the computations, we make the
system isotropic in the $yz$-plane by the rescaling
$z'=z{(c_{66}/c_{44})}^{1/2}$ and $u'(\br')=u(\br)$. In the following
we will omit the primes.

The critical droplet is a solution of the saddle point equation for
the Hamiltonian of Eq.~(\ref{eq:effectiveH}) with fixed boundary
conditions $u(0,\rho\to0)=(n+1)\ell$ and $u(0,\rho\to\infty)=n\ell$
with $\rho=(y^2+z^2)^{1/2}$ and $n$ integer. For the precise solution
see, e.g., a related discussion for a single strong impurity in a
Luttinger liquid by \citet{Giamarchi}. The shape of the droplet
  is characterized by its radius $R$ and the width $w$ of the droplet
  wall so that the displacement field obeys approximately 
\begin{align}\label{eq:droplet}
u(0,\rho)\approx \left\{ \begin{array}{ll}
(n+1)\ell, & \rho\in [0,R]\\
n\ell, & \rho\in [R+w,\infty].
\end{array} \right.
\end{align}
The exact shape of the
droplet wall is not essential for the discussion that follows. We
assume that it smoothly interpolates between $(n+1)\ell$ and
$n\ell$. The width $w$ of the droplet wall does not depend on the
radius for small currents $J\to 0$. In this limit the critical droplet
radius is much larger than the width $R\gg w$ so that the energy loss
is balanced by the energy gain from the Lorentz force (see
e.g.\cite{Malinin+04} and references therein). The critical droplet
radius and energy is determined by maximizing the droplet energy
$E_{drop}(R)$.

Since across the droplet wall the FLs are not aligned to the defect
plane, a strong defect tends to reduce the width of the wall. Then,
the large $\bq$ behavior of the propagator of
Eq.~(\ref{eq:effectivepropagator}),
$[\widetilde{\mathcal{G}}(0,\bq)]^{-1}\approx (\pi c_{66}
q^2)/\Lambda$, becomes important at the wall since it describes
elastic deformations on small length scales. For sufficiently strong
defects ($w\simeq \Lambda^{-1}$) the precise form of the droplet wall
is determined by the interplay between the elastic energy $\sim q^2$
and the defect pinning energy. The energy loss for FLs at the defect plane is \cite{Giamarchi} $E_{core}\sim R v^{1/2}$, as
known from droplets in the sine-Gordon model.

The elastic energy loss outside the plane, due to the deformation
of the FLs at the defect plane is determined by the low $\bq$ behavior
of the propagator of Eq.~(\ref{eq:effectivepropagator}),
$[\widetilde{\mathcal{G}}(0,\bq)]^{-1}\approx
2\sqrt{c_{11}c_{66}}|\bq|$, since the deformation occurs across
  the large scale $R$. It captures the three-dimensional nature of
the FLL by its non-local form $\sim |\bq|$. This
is obvious when the elastic energy is written as
\begin{align} \label{eq:real}
\cH_{el}=\frac{\sqrt{c_{11}c_{44}}}{4\pi}\int\dif^2 \br_1\int\dif^2 \br_2\frac{[u(\br_1)-u(\br_2)]^2}{(\br_1-\br_2)^3} \, .
\end{align}
Here $\br_1$ and $\br_2$ lay in the defect plane and satisfy
$|\br_1-\br_2|>\lambda$.  The long ranged elasticity in the effective
two-dimensional elastic Hamiltonian of Eq.~(\ref{eq:real}) results
from fluctuations outside the defect plane that have been integrated
out. Since $R\gg w$, the precise form of the droplet wall is not
important for estimating $\cH_{el}$ and we can assume $w=0$ (corresponding
to $v\to \infty$). Then we obtain for the elastic energy
\begin{align}\label{eq:interaction_kinks}
E_{el}\approx 2\sqrt{c_{11}c_{44}}\ell^2R\log{(R/\lambda)} \, .
\end{align}
This result can be interpreted as the energy of charges of equal sign
(corresponding to kinks in the displacement field) which are
  placed along a circle of radius $R$ and interact via the
  three-dimensional Coulomb potential.  

The energy losses mentioned above are balanced by an energy gain
  due to the Lorentz force that is described by the last term of
Eq.~(\ref{eq:effectiveH}),
\begin{align}\label{eq:gain}
E_{force}\approx\ell R^2\pi \sqrt{\frac{c_{44}}{c_{66}}}\int \dif x {J(x)B(x)} \, .
\end{align}
When we estimate $E_{force}$, a finite width of the droplet wall can
be neglected since $R\gg w$. The total droplet energy then reads
\begin{align}\label{eq:dropletsingle}
E_{drop}&\approx E_{el}+E_{core}-E_{force}\notag\\
&\approx\alpha R\log\left(\frac{R}{\lambda}\right)+R\beta-\gamma R^2 \, ,
\end{align}
where $\alpha=2{(c_{11}c_{44})}^{1/2}\ell^2$, $\beta\sim v^{1/2}$ and
$\gamma=\pi {(c_{44}/c_{66})}^{1/2}\ell\int \dif x{J(x)B(x)}$.
The creep rate is
  determined by the droplet with the largest total energy $E_{drop}$
  which is called the critical droplet. Solving the equations
$\partial_R E_{drop}=0$ and $\partial_R^2 E_{drop}<0$ we find the size
$R^*$ and the energy $E_{drop}^*$ of the critical droplet.  Increasing
of the droplet radius beyond $R^*$ does not cost any energy and
droplet freely expands. For $R\to\infty$ the system reaches a new
metastable state in which all FLs are shifted by $\ell$
perpendicular to the defect. The nonlinear resistivity is given by the
Arrhenius law. In the limit of a vanishing current density, i.e., for
large $\beta/\gamma$, we get
\begin{align}\label{eq:creep}
\rho&\sim e^{-E_{drop}^*/T}\notag\\&\sim \exp{\left\{-
\frac{1}{4\gamma T }\left[\left(\beta+\alpha\log{\frac{\beta+\alpha}{2\lambda \gamma }}\right)^2-\alpha^2\right]\right\}}.
\end{align}
Prefactors are not determined here
since in the limit of a vanishing current density the
  current-voltage characteristic is dominated by the exponential
  factor of Eq.~(\ref{eq:creep}).  The result for $\rho$ yields the
creep exponent $\mu=1$ plus logarithmic corrections.  To estimate the
coefficient $\gamma$ we need to know how the current density and the
magnetic induction vary in space. This is a tedious analysis which
goes beyond the scope of the present study and is left for further
investigation.

Next we examine how randomly distributed weak point impurities around
the defect plane affect the FL creep for $J\to 0$. A weak defect is
relevant in the RG sense only for $g<1$ and it then flows to the
strong coupling limit. Criteria for the relevance of a strong
  defect with an arbitrary orientation are not available. Therefore
we study below only the strong coupling limit for a defect that is
oriented parallel to the main FLL planes. We expect that the
liberation from the defect plane is the limiting factor for the FL
motion so that the creep exponent is reduced compared to its value in
the BG phase.

The shape of the droplet is again given by Eq.~(\ref{eq:droplet}) but
now the impurities control the fluctuations of the FLs outside the
defect. Without point impurities, the displacement field decays
outside the defect plane as $u(x,\rho=0)\approx (\ell/2) (R/x)^2$ for
$x\gg R$. Point-like impurities induce additional displacement
fluctuations and the droplet-induced deformations are no longer
 long-ranged. This can be seen from the correlation function
in the presence of a relevant defect that follows from
Eq.~(\ref{eq:correlationrelevant}),
\begin{align}\label{eq:impurities}
  \Big\langle[u(x,\rho)-&u(0,\rho)]^2\Big\rangle=\notag\\&\frac{1}{2}\Big
  \langle[u(x,\rho)-u(-x,\rho)]^2\Big\rangle_{BG} \, .
\end{align}
The subscript BG means that the correlation function is computed at
the BG fixed point. Since the right hand side of
Eq.~(\ref{eq:impurities}) is of the order $a^2$ for $x\approx L_a$ we
conclude that the droplet extends up to $L_a$ from the defect. This
yields for the energy gain from the Lorentz force the rough estimate
$E_{force}\approx JBL_aR^2\ell$ where $J$ is the mean current
density inside the droplet. Hence, in the limit $J\to 0$ resistivity
is
\begin{align}
  \rho(J)\sim
  \exp{\left[-\frac{C_1}{J}\left(\log{\frac{C_2}{J}}\right)^2\right]}
  \, ,
\end{align}
where $C_1$ and $C_2$ depend on $T$, $B$, on the strength and
concentration of the impurities, and on the defect strength. This
result shows that a single relevant defect plane indeed slows down the
FL creep in comparison to the BG phase.
\subsection{Finite density of weak defects}
\label{section:creep_weak}
Here we consider FL creep perpendicular to many weak defects with
  random positions but in the absence of point impurities,
cf.~Sec.~\ref{section:weak}. The motion of FL bundles under the
influence of the Lorentz force is again driven by the nucleation of
critical droplets\cite{Blatter+94}. A typical droplet is schematically shown in Fig.~\ref{Fig:3}. For small currents, the droplet extends
over many defect planes in order to balance elastic and pinning energy
loss with bulk energy gain from the Lorentz force. For small current
densities, the FLL is properly described in terms of the interface
tensions $\Sigma_y$ and $\Sigma_z$, see Eq.~(\ref{eq:newterm}), which
are appropriate on sufficiently large length scales.  The energy of
the droplet is then of the form
\begin{align}\label{eq:E-nucl}
E_{drop}\approx\sqrt{\frac{c_{44}}{c_{66}}} L_x^{d-2} R^2 \left(\frac{c_{11}\ell^2}{L_x^2}+\frac{\Sigma_y}{R}-JB\ell \right) \, .
\end{align}
The elastic energy cost for the formation of the droplet consists of
two terms. The first term of Eq.~(\ref{eq:E-nucl}) is the energy of a
wide domain wall of width $\sim L_x$ parallel to
$yz$-plane. The second term of Eq.~(\ref{eq:E-nucl}) describes the
energy of a narrow cylindrically shaped domain wall perpendicular to
$yz$-plane. In the estimate of $E_{drop}$ we have taken into account
that the elastic energy and the energy from planar disorder scale
in the same way. The last term of Eq.~(\ref{eq:E-nucl}) is the energy
gain from the Lorentz force. $J$ ($B$) is to be understood as the mean
current density (magnetic induction) averaged over the defect spacing.

Note that we have used again the rescaling
$z'=(c_{66}/c_{44})^{1/2} z$. In Eq.~({\ref{eq:E-nucl}}) we have
taken into account the logarithmic roughness of the displacement
field, corresponding to the roughness exponent $\zeta=0$. We have
ignored logarithmic corrections. The $\sigma$-term of
Eq.~(\ref{eq:replicaHam}), that is responsible for the roughness
exponent $\zeta=1/2$, can be eliminated by the simple transformation
that was discussed in Sec.~\ref{section:weak} and hence it does not
affect the FL dynamics. In case of a potential
  breakdown of the $\epsilon$ expansion (cf.~Sec.~\ref{section:FRG})
  in $d=3$ and the existence of a strong coupling fixed point
  \cite{ViFE84} see Sec.~\ref{section:creep_strong}.

To determine the critical droplet we solve $\partial_{L_x}E_{drop}=0$
and $\partial_{R}E_{drop}=0$.  We find for the critical radius $R^*$
and the critical length $L_x^*$ of the droplet
\begin{align}
  \frac{R^*}{\Sigma_y}\sim
  \frac{L_x^{*2}}{c_{11}\ell^2}\sim \frac{1}{JB\ell} \, .
\end{align}
This yields for $d=3$ the nonlinear resistivity in the limit $J\ll
J_D$,
\begin{equation}\label{eq:rho(J)}
 \rho(J)\sim
 e^{-\left({J_D}/{J}\right)^{3/2}},\,\,\,\,J_D
 ={\cal C}\frac{(\Sigma_y\Sigma_z)^{2/3}( c_{11}/\ell)^{1/3}}{B T^{2/3}}.
\end{equation}
Here $\cal C$ is positive numerical constant of order unity. Thus the
non-linear resistivity is considerably reduced compared to the Bragg
glass phase and to a single defect plane in the presence of impurities.

In the similar way one can consider the creep in the presence of
randomly distributed columnar defects that are aligned to the applied
magnetic field and have a mean spacing that is larger than FL spacing.
Based on a functional RG in $d=5-\epsilon$ dimensions\cite{Balents},
we obtain the creep exponent $\mu=1$. This result is in agreement with
the one derived in \citet{Blatter+94} by other means. The result is
expected to apply to the weakly pinned Bose glass phase.

\subsection{Finite density of strong defects}
\label{section:creep_strong}
Here we discuss the analog of the previous subsection in the limit of
strong defects, see Sec.~{\ref{section:strong}}.  The ground state
degeneracy given by Eq.~(\ref{eq:ground}) is broken when a current is
applied. The system evolves between different ground states via the
formation of critical droplets.  For small currents $J\ll {c_{11}
\ell}/{(B \ell_D^2)}$ the critical droplet extends over many defect
planes and we obtain the creep exponent $\mu=3/2$. For moderate
currents with ${c_{11}\ell}/{(B \ell_D^2)}\ll J\ll {v}/{(\phi_0\xi
  \ell_D)}$ the droplet forms only at a single defect plane and we
recover Eq.~(\ref{eq:creep}) with
$\gamma\approx{(c_{44}/c_{66})}^{1/2}\ell\ell_D{JB}$, i.e., a
creep exponent $\mu=1$.

We assume that the saddle point solution for $u_i$ with fixed boundary
conditions $u_i(\rho\to 0)={u}^{(n+1)}_i$ and
$u_i(\rho\to\infty)={u}^{(n)}_i$ obeys Eq.~(\ref{eq:droplet}). At each defect plane, the radius
$R_i$ and the center of the droplet can be different. However, it is
plausible to assume that in the saddle point configuration the droplet
is centered at the same lateral position in each plane and all
droplets have the same radius $R$ since the droplet tends to maximize
its volume while keeping the surface minimal.  Specifically, we assume
that the droplet is located between the $s$th and the $(s+m)$th defect
plane.

The width of the droplet wall for a sufficiently strong defect
potential satisfies $w\ll \ell_D$ and $w\ll R$. The precise form of
the wall is not important in finding the energy of a domain wall
parallel to the defects as well as the bulk energy gain. Hence, we set
the width of the droplet to zero which yields for the Fourier transform of
the displacement at the $i$th defect plane
\begin{align}\label{eq:solution}
\widetilde{u}_i(\bq)=2 \pi R \ell\frac{J_1(qR)}{q}+u_i^{(n)}
 (2\pi)^2\delta(\bq) \, ,
 \end{align}
 where $J_1$ is the Bessel function of the first kind.  Due to the
 transport current we have to add to the Hamiltonian of
 Eq.~(\ref{eq:reducedH}) the additional energy
\begin{align}\label{eq:H_force}
H_{force}=-\sum_{i=1}^{N_D-1}\int_{x_i}^{x_{i+1}}&\dif x f(x)\Big\{ \frac{\widetilde{u}_i(\mathbf{0})}{\Delta x_i}(x_{i+1}-x)\notag\\& +\frac{\widetilde{u}_{i+1}(\mathbf{0})}{\Delta x_i}(x-x_{i})\Big\} \, ,
\end{align}
where $f(x)=(c_{44}/c_{66})^{1/2}J(x)B(x)$. By substituting the
Eq.~(\ref{eq:solution}) into Eq.~(\ref{eq:reducedH}), we find that the
droplet energy has a different form for $R\gg \ell_D
\sqrt{c_{66}/c_{11}}$ and $R\ll \ell_D \sqrt{c_{66}/c_{11}}$.

\begin{figure}
\includegraphics[width=0.8\linewidth]{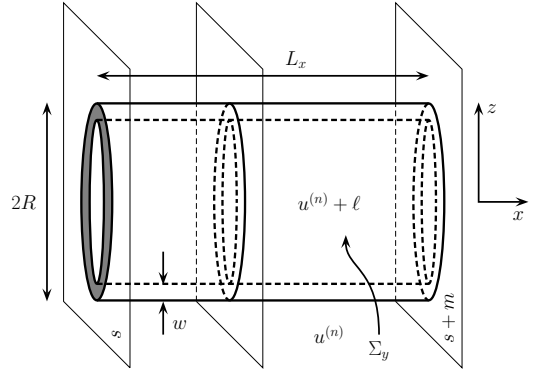}
\caption{Schematic representation of a cylindrically shaped droplet
of radius $R$ and length $L_x$ that extends across more than one
defect plane. It drives the system into the new metastable state
$u^{(n)}+\ell$. Note that domain walls parallel to the defects are wide (not shown here), while the cylindrically shaped domain wall is narrow (of width $w$).}\label{Fig:3}
\end{figure}

First we discuss the case $R\gg \ell_D \sqrt{c_{66}/c_{11}}$. The
droplet energy reads then 
\begin{align}\label{eq:droplet energy}
E_{drop}=\Sigma_x R^2 \pi+\Sigma_y 2 R \pi m \ell_D-{JB}\ell
R^2\pi m\ell_D \sqrt{\frac{c_{44}}{c_{66}}}\, .
\end{align}
In order to explain and interpret the first term of Eq.~
(\ref{eq:droplet energy}), we consider an excited state with
\begin{align}
u_i^{\pm}=\left\{ \begin{array}{ll}
u_i^{(n)} & \text{for}\quad i\le k \\
u_i^{(n\pm1)} &  \text{for}\quad i>k \, .
\end{array} \right.
\end{align}
This state describes a domain wall parallel to the defects. Using
Eq.~(\ref{eq:ground_hamiltonian}), it can be shown that for a given
disorder realization the energy cost of such wall per unit surface
area is
\begin{align}
\Sigma_x^{\pm}(k)=c_{11}\sqrt{\frac{c_{44}}{c_{66}}}\frac{\ell^2}{2\Delta x_k}\left[1\pm
2\left(\frac{\Delta \alpha_k}{\ell}-
\left[\frac{\Delta
\alpha_k}{\ell}\right]_G\right)\right] \, .
\end{align}
Since the droplet consists of two such walls (see Fig.~\ref{Fig:3}),
the surface tension in the first term of Eq.~(\ref{eq:droplet
energy}) is given by
\begin{align}
\Sigma_x=\Sigma_x^{+}(s-1)+\Sigma_x^{-}(s+m) \, .
\end{align}
The droplet takes advantage of fluctuations in the surface tension
$\Sigma_x$ and the lowest value of $\Sigma_x$ for a droplet of length
$L_x=m\ell_D$ is typically of the order
$c_{11}\sqrt{{c_{44}}/{c_{66}}}\ell^2/L_x$ when $\ell_D\gg \ell$. We
point out that this result is in agreement with the statistical tilt
symmetry.

The second term of Eq.~(\ref{eq:droplet energy}) is the energy cost
for the domain wall perpendicular to the defects with surface tension
 $\Sigma_y$. We do not provide here an explicit expression for
 $\Sigma_y$, for the reasons discussed in
 Sec.~\ref{section:strong}. Note that $\Sigma_y$ carries information
 about the strength and density of defect planes. By ignoring the spatial variations of $J$ and $B$, we get the
average bulk energy gain to be given by the last term in Eq.~(\ref{eq:droplet energy}).
 
By solving $\partial_m E_{drop}=0$ and $\partial_R E_{drop}=0$ we
determine the radius $R^*$ and length $L_x^*=m^*\ell_D$ of the
critical droplet. With this parameters, Eq.~(\ref{eq:droplet energy})
yields the energy of the critical droplet. We find that this energy
leads, up to unimportant prefactors, to the same nonlinear
resistivity as in the weak pinning case of Eq.~(\ref{eq:rho(J)}).
This result is valid for sufficiently small currents such that $R^*\gg
\ell_D \sqrt{{c_{66}}/{c_{11}}}$ and $m^*\gg 1$. The latter condition
results from treating $m$ as a continuous variable which is a
reasonable approximation for critical droplet that extend across a
large number of defect planes. These conditions translate to the
requirement $J\ll {c_{11} \ell}/{(B \ell_D^2)}$ for the current
density.

Next, we shortly discuss the droplet expansion when it reaches the
radius $R^*$ and the length $L_x^*$. By analyzing the Hessian matrix
of $E_{drop}$ we find that the eigenvector that points
into the direction of the free droplet expansion
has a $x$-component that is much smaller than its $\rho$-component, i.e.,
an expansion of the cross-section of the cylinder is favorable over an expansion of its length.
In order to describe a potential growth in the $x$-direction,
 one has to know the set of numbers
$\Sigma_x(i)^{\pm}$ that depends on the disorder realization. However,
the droplet will get stuck between planes with low $\Sigma_x$ and a further
expansion along the $x$-axis costs energy. 

Droplets occur and expand independently across the
entire sample. After the FLs have moved in some regions, the
boundaries of these regions will be favorable sites for emergence of
new droplets\cite{Malinin+04}. Indeed, after the formation and
expansion of a droplet up to the $(i-1)$th
defect plane such that $u_k=u_k^{(n+1)}$ for $k\le i-1$ and
$u_k=u_k^{(n)}$ for $k\ge i$, the new surface tension reads
\begin{align}
\Sigma_x^{'+}(i)&\approx c_{11}\sqrt{\frac{c_{44}}{c_{66}}}\frac{1}{\ell_D}\left[-1+
2\left(\frac{\Delta\alpha_{i}}{2\pi}-\left[\frac{\Delta\alpha_{i}}{2\pi}
\right]_G\right)\right]\notag\\&=-\Sigma_x^{-}(i)<0 \, .
\end{align}
The reason for this result is that after the droplet
expansion an additional FLL plane appears with respect to the initial
ground state between $i$th and $(i-1)$th planes and the FLs are
compressed. The formation of a new droplet at the
$i$th plane is favorable because it allows the system to relax into
the new ground state configuration between $(i-1)$th
and $i$th plane. Since the droplet described by Eq.~(\ref{eq:droplet energy}) has the longest life time, the resistivity is determined by 
$\rho\sim \exp{\left(-{E_{drop}^*}/{T}\right)}$, where the critical energy
$E_{drop}^*$ is given by Eq.~(\ref{eq:droplet energy}) evaluated at
$R^*$, $L_x^*$.

When comparing Eq.~(\ref{eq:rho(J)}) with the creep exponent $\mu=1/2$
of the defect free BG phase, we see that defect planes act as a more
efficient source of pinning in stabilizing superconductivity than
point impurities. However, we have considered only typical droplets in
estimating $\Sigma_x$. For system sizes $L\gg \ell_D$ it is likely
that rare regions with untypically large $\Sigma_x$ will appear and in
turn determine the resistivity. We leave this problem for further
investigations.

Next, we consider the case $R\ll\ell_D\sqrt{c_{66}/c_{11}}$. From the
second term of Eq.~(\ref{eq:reducedH}) we find that for
$m$ defect planes the energy loss given by
Eq.~(\ref{eq:interaction_kinks}). The first term in
Eq.~(\ref{eq:reducedH}), that describes the coupling between
neighboring defects, provides a much smaller contribution than
  the second term and can be neglected. Then the defects are
effectively decoupled, and the nucleus energy is
\begin{align}
E_{drop}=mE_{single}\, .
\end{align}
Here $E_{single}$ is the energy of the droplet that appears in the
case of a single defect plane. It is given by
Eq.~(\ref{eq:dropletsingle}) with the system size replaced by the mean
defect distance $\ell_D$. The nucleus energy grows with increasing $m$
so that it is minimal for $m=1$ and the critical droplet is located at
one defect plane only. Nonliner resistivity is then again given by
Eq.~(\ref{eq:creep}), but with
$\gamma\approx{(c_{44}/c_{66})}^{1/2}\ell\ell_D{JB}$. This result
is valid for intermediate currents ${c_{11} \ell}/{(B \ell_D^2)}\ll
J\ll {v}/{(\phi_0\xi \ell_D)}$.

\section{Discussion}
\label{section:conclusions}
\begin{figure}
\includegraphics[angle=90,width=0.9\linewidth]{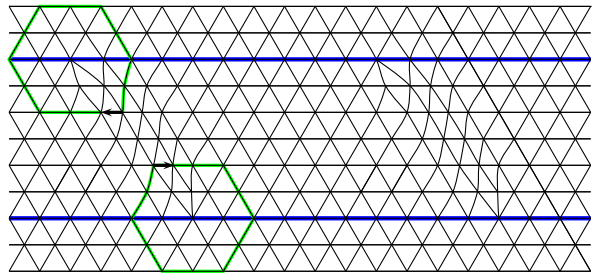}
\caption{(Color online) Array of edge dislocations that is located at
  the defects (thick blue lines) in order to relax shear strain. The
  Burgers vector is parallel to the defects.}\label{Fig:dislocation}
\end{figure}

For a finite density of randomly distributed parallel planar defects
with the magnetic field aligned to them, and with a mean defect
spacing that is larger than FL spacing, we find a new phase of FLs at
low temperatures, the planar glass.  We considered mainly a simplified
model with an uniaxial displacement field (which is also applicable
to a wide class of other systems). Here we comment on possible
consequences of this simplification by taking into account also the
displacement field $u_y$ parallel to the defects. The part of the
defect Hamiltonian that describes the coupling of the defect potential
to the slowly varying part of FL density can be eliminated by
transforming $u_x$ as described by Eq.~(\ref{eq:transformation}).  For
strong planar defects each defect plane is occupied by a single FL
layer and hence $u_x^{(n_i)}(x_i,y,z)=\ell n_i+\alpha_i$ for all $y,z$
in order to maximize the pinning energy gain.  Even in the absence of
point disorder the displacement $u_y$ does not vanish. This can be
seen most easily in the case of isotropic elasticity where the
following relations hold
\begin{align}
 \sigma\partial_x u_x =-
\partial_y u_y,\qquad \sigma=\frac{c_{11}-c_{66}}{c_{11}+c_{66}}\, .
\end{align}
Here $\sigma$ is the Poisson number with $-1<\sigma<1$.
\cite{Landau7} The strain $\partial_x u_x$ in the gap between the
defects at $x_{i+1}$ and $x_i$ is
\begin{align}
\partial_x u_x\approx
1+\frac{v\rho_0}{c_{11}\Delta x_i}+\ell\frac{\Delta n_{i}}{\Delta x_{i}}\, ,
\end{align}
where we used the notation $\Delta A_i=A_{i+1}-A_i$ for a variable
$A$.  The difference of the strain $\partial_y u_y$ in neighboring
gaps is then
\begin{align}
 \Delta\partial_y u_y\approx -\sigma\ell\left[\frac{\Delta n_{i+1}}{\Delta x_{i+1}}-\frac{\Delta n_i}{\Delta x_i}+\frac{v\rho_0}{c_{11}\ell}\left(\frac{1}{\Delta x_{i+1}}-\frac{1}{\Delta x_{i}}\right)\right]
\end{align}
which is of the order $\pm \sigma \ell/\ell_D$. On the scale $L_y$
this implies $\Delta u_y\sim \pm \sigma\ell L_y/ \ell_D$. To avoid a
diverging shear energy, dislocations with a Burgers vector parallel to
the $y$-direction occur at the defect planes (see
Fig.~\ref{Fig:dislocation}).  Their distance along the $y$-direction
is of the order $\ell_D/\sigma$. The energy of a pair of edge
dislocations with anti-parallel Burgers vectors at a distance $\ell_D$
is \cite{Br}
\begin{align}
E_{\emph{edge}}\approx
\frac{c_{66}}{2\pi}a^2L_z\ln\left(\frac{\ell_D}{a}\right) \, .
\end{align}
This energy has to be compared with the energy gain from the defects which
is of the order
\begin{align}
 E_{D}\approx-\frac{\ell}{\sigma\xi}L_zv\ell_D\rho_0 \, .
\end{align}
Hence for $\sigma c_{66}a^3\xi \ll \ell_D v$ the energy of the
dislocations is overcompensated by the defect planes and dislocations
will be present.

In general, the network of additional FLL sheets spanned by the
dislocations will be complicated. The network follows from the
solution of the equations of two-dimensional elasticity with the
boundary condition $u_x(x_i,y,z)\equiv u_x^{(n_i)}(x_i,y,z)$ and the
dislocation density $b_y(x_i,y)$ at each defect. The energy has to be
minimized first with respect to $b_y(x_i,y)$ and the with respect to
$n_i$ \cite{ArNa01}. The resulting state is completely ordered along
the $z$-direction. It is also ordered in the sense that the interface
tensions $\Sigma_z$ and $\Sigma_y$ are non-zero.  A change in the
boundary conditions with $u_x(x,y,z\to\infty)=u_x(x,y,z\to 0)+\ell$
increases again the energy. Hence the transverse Meissner effect as
well as the resistance against FLL shearing perpendicular to the
defects are still present. Bond-orientational order\cite{Halperin+78}
persists since disclination pairs remain bounded in the cores of the
edge dislocations.

Since the Burgers vectors of the dislocations are always parallel to
the defect planes, creep along the $x$-direction is not
facilitated. Under the assumption that the distribution of $\Sigma_x$
is uniform even in the presence of dislocations we recover the creep
law of Eq.~(\ref{eq:rho(J)}). To describe creep parallel to the
defects one has to take into account the interaction between
dislocation, a problem not considered so far
\cite{Kierfeld+00,Zaiser+04}.  We leave this for further studies.  For
weak pinning qualitatively the same behavior can be expected on scales
$L_x\gg L_D$, in particular if the defect potential flows under
  the RG to strong coupling. If the sample exhibits two orthogonal
families of (non-intersecting) defects, long range order in the $x$
and $y$ direction is destroyed even without point impurities on scales
larger than $L_D$. The creep is then limited by the slowest mechanism
and hence Eq.~(\ref{eq:rho(J)}) is likely to be valid for all current
directions in the $xy$-plane.

\section*{Acknowledgments}
The authors acknowledge helpful comments from F. de la Cruz, J.
Kierfeld, D.~R. Nelson, L. Radzihovsky, Z. Ristivojevic, V.~M. Vinokur, R.
Woerdenweber, E. Zeldov and M. Zaiser. Financial support by the Deutsche Forschungsgemeinschaft through Sonderforschungsbereich 608 (AP and TN)
and through the Heisenberg Program under grant No.~EM70/3 (TE) is acknowledged.

\appendix
\section{Replica Hamiltonian for the defect free system}

In this Appendix the Hamiltonian for defect-free system is derived,
using the replica approach for averaging over point impurities. The
pinning energy of randomly distributed impurities reads (see
Sec.~\ref{section:BG})
\begin{align}\label{eq:Hpinning}
\cH_{P}=\int d^3\br\;V_{P}(\br)\rho_0\left\{- \bm{\nabla}_{\bx} \bu(\br) +
  \sum_{\bG\neq 0}e^{i\bG[\bx-\bu(\br)]} \right\},
\end{align}
where $\bx=(x,y)$. If the system is characterized by a roughness
exponent $\zeta$, displacements vary with the scale $L$ as $u\sim
L^{\zeta}$, $\zeta<1$. The elastic energy scales as $L^{d-2+2\zeta}$, and the
first and second term of Eq.~(\ref{eq:Hpinning}) scale as
$L^{(d-2+2\zeta)/2}$ and $L^{d/2}$, respectively.  These simple
scaling arguments  show that the coupling of the divergence of the
displacement field to the disorder potential is irrelevant with
respect to the elastic energy in $d>2$ and the second term of
Eq.~(\ref{eq:Hpinning}) is relevant for $d<4$. Since we are interested
in the behavior on large length scales in $d=3$ dimensions, we can neglect the
first term of Eq.~(\ref{eq:Hpinning}).  After performing the disorder average, the replicated pinning energy reads
\begin{align}\label{replica}
\cH_{P}^n&\approx-\frac{v_p^2n_{imp}\rho_0^2}{2T}\sum_{\alpha,\beta=1}^n\int_{\br,\br'}\delta_{\xi}(\bx-\bx')\delta(z-z')\notag\\&\sum_{\bG,\bG'\neq0}
e^{i\bG[\bx-\bu^{\alpha}(\br)]+i\bG'[\bx'-\bu^{\beta}(\br')]}\notag\\&=- \frac{v_p^2n_{imp}\rho_0^2}{2T}\sum_{\alpha,\beta=1}^n\int_{z,\bx}\sum_{\bG\neq0} e^{i\bG[\bx-\bu^{\alpha}(\bx,z)]}\notag\\& \sum_{\bG'\neq 0}\int_{\bx_r}\delta_{\xi}(\bx_r)e^{i\bG'[\bx+\bx_r-\bu^{\beta}(\bx+\bx_r,z)]}\notag\\ 
&\approx-\frac{v_p^2n_{imp}\rho_0^2}{2T}\sum_{\alpha,\beta=1}^n\int_{z,\bx}\sum_{\bG,\bG'\neq0} e^{i\bx(\bG+\bG')}\notag\\& e^{-i[\bG \bu^{\alpha}(\bx,z)+\bG'\bu^{\beta}(\bx,z)]} {\delta}_{\xi^{-1}}(\bG').
\end{align}
Using the relative coordinate $\bx_r =\bx'-\bx$ and the fact that
$\delta_{\xi}(\bx_r)$ is nonzero only for $|\bx_r|\leq\xi$, we
approximate the slowly varying displacement field as
$\bu^{\beta}(\bx+\bx_r,z)\approx\bu^{\beta}(\bx,z)$, where
${\delta}_{\xi^{-1}}(\bG)$ is the Fourier transform of 
$\delta_{\xi}(\bx)$.  Since the displacement field varies slowly on
the scale of the FLL constant, the integral over $\bx$ vanishes for all
combinations of $\bG$ and $\bG'$ except for $\bG=-\bG'$ when
the oscillatory factor $e^{i\bx(\bG+\bG')}$ becomes one. The replicated pinning
Hamiltonian can now be written as
\begin{align}
\cH_{P}^n&=-\frac{v_p^2n_{imp}\rho_0^2}{2T}\sum_{\alpha,\beta=1}^n\int_{\br}\sum_{\bG\neq0} e^{i\bG[\bu^{\alpha}(\br)-\bu^{\beta}(\br)]}{\delta}_{\xi^{-1}}(\bG)\notag\\&=
-\frac{1}{2T}\sum_{\alpha,\beta=1}^n\int_{\br}R_P[\bu^{\alpha}(\br)-\bu^{\beta}(\br)].
\end{align}

\section{Effective Hamiltonian on the defect plane}
\label{app:effective}

In this appendix we present a functional integral approach for the
derivation of the effective Hamiltonian on the defect plane for the
case of a FLL with point impurities and a single defect plane.  Since
the pinning energy of the planar defect involves only the displacement
perpendicular to the defect, $u_x$, we integrate out $u_x$ outside the
defect and $u_y$ across the entire sample. The partition function can
be written as
\begin{align}
Z&=\int\mathcal{D}\varphi(\br_D)\int\mathcal{D}\bu(\br)e^{-\frac{\cH}{T}} \prod_{\br_D}\delta[u_{x}(\br_D)-\varphi(\br_D)],
\end{align}
where $\cH=\cH_0+\cH_P+\cH_D$.  The displacement at the defect is
constrained to be $u_x(\br_D)=\varphi(\br_D)$ and this constraint is
implemented by $\delta$-functions in the functional integral. The
defect is aligned to the magnetic field and $\br_D$ is given by
Eq.~(\ref{eq:defect-para}). After averaging over point impurities we
get
\begin{widetext}\begin{align}
\overline{Z^{n}}&=\int\mathcal{D}[\varphi^{\alpha}(\br_D)]e^{-\sum_{\alpha} \frac{\cH_D(\varphi^{\alpha})}{T}}\int \mathcal{D}[\bu^{\alpha}(\br)]\;
e^{-\frac{\cH^{n}_0}{T}}
\prod_{\alpha,\br_D}\delta[u_{x}^{\alpha}(\br_D)-\varphi^{\alpha}(\br_D)]\notag\\
&=\int\mathcal{D}[\varphi^{\alpha}(\br_D)]e^{-\frac{\cH_{eff}^{n}(\varphi^{\alpha})}{T}},
\end{align}
where $\cH_0^n$ is given by Eq.~(\ref{eq:Replica Hamiltonian
  quadratic}) and $\int\mathcal{D}[\varphi^{\alpha}(\br_D)]=\prod_{\alpha=1}^{n}\int \mathcal{D}\varphi^{\alpha}(\br_D)$.  Using the function integral representation
\begin{align}\label{eq:dirac delta}
\prod_{\alpha,
\br_D}\delta[u_{x}^{\alpha}(\br_D)-\varphi^{\alpha}(\br_D)]=\int
\mathcal{D}[\Lambda^{\alpha}(\br_D)]\;e^{i\sum_{\alpha}\int_{\br_D}
\Lambda^{\alpha}(\br_D)[u_{x}^{\alpha}(\br_D)-\varphi^{\alpha}(\br_D)]}
\end{align}
of the $\delta$-function, we obtain for the effective
replica Hamiltonian $\cH_{eff}^{n}$ the equation
\begin{align}
e^{-\frac{\cH_{eff}^{n}}{T}}=e^{-\sum_{\alpha}\frac{\cH_{D}(\varphi^{\alpha})}{T}}\;\int\mathcal{D}[\Lambda^{\alpha}(\br_D)] \left\langle
e^{i\sum_{\alpha}\int_{\br_D}
\Lambda^{\alpha}(\br_D)[u_{x}^{\alpha}(\br_D)-\varphi^{\alpha}(\br_D)]}
\right\rangle_{\cH^n_0}
\end{align}
up to a constant. Here $\langle \ldots\rangle_{\cH}$ denotes the
thermal average with respect to $\cH$. For this average we obtain
\begin{align}
\left\langle e^{i\sum_{\alpha}\int_{\br_D}
\Lambda^{\alpha}(\br_D)[u_{x}^{\alpha}(\br_D)-\varphi^{\alpha}(\br_D)]}
\right\rangle_{\cH^n_0}=e^{-i\sum_{\alpha}\int_{\br_D}
\Lambda^{\alpha}(\br_D) \varphi^{\alpha}(\br_D)}\;
e^{{-\sum_{\alpha,\beta}\frac{1}{2}\int_{\br_{D1},\br_{D2}}
\Lambda^{\alpha}(\br_{D1})\; \Lambda^{\beta}(\br_{D2})\; T
\;\mathcal{G}_{xx}^{\alpha,\beta}(\br_{D1}-\br_{D2})}},
\end{align}
where $\langle \widetilde{u}_{x}^{\alpha}(\bq)\widetilde{u}_{x}^{\beta}(\bq')\rangle_{\cH_0^n}=T (2\pi)^{d}
\widetilde{\mathcal{G}}_{xx}^{\alpha,\beta}(\bq) \delta(\bq+\bq')$.
The effective Hamiltonian reads
\begin{align}
\cH_{eff}^n=-\sum_{\alpha}\sum_{k>0} 2 v_k \rho_0 \int d\br_D\;
\cos{\{kG_D[\delta-\varphi^{\alpha}(\br_D)]\}}
+\frac{1}{2}\sum_{\alpha,\beta} \frac{1}{(2\pi)^{d-1}}\int
\;d^{d-1}\;\bq\; \varphi^{\alpha}(\bq)\;
\widetilde{\boldsymbol{\mathcal{Q}}}^{-1}_{\alpha,
\beta}(\bq)\;\varphi^{\beta}(-\bq),
\end{align}
where
$\widetilde{\boldsymbol{\mathcal{Q}}}(\bq)=\widetilde{\mathcal{G}}_{xx}(x=0,\bq)$
and here $\bq$ is the in-plane momentum.

\section{Density oscillations in the presence of an irrelevant defect plane}
\label{app irrelevant}

In this appendix we analyze the thermal and disorder average of the FL
density around an irrelevant defect plane in the presence of point
impurities by perturbation theory in $v$. For the local density
variations we have
\begin{align}
\overline{\langle\delta\rho[\br,\bu(\br)]\rangle}=\lim_{n\to 0}\int \mathcal{D}[\bu^{\alpha}]\; \delta\rho[\br,\bu_1(\br)]\; e^{-\beta\left[\cH_0^n+\sum_{\alpha}\cH_D(u^{\alpha}_x)\right]},
\end{align}
where $\delta\rho[\br,\bu(\br)]=\rho[\br,\bu(\br)]-\rho_0$ and
$u_1(\br)$ is the displacement field with replica index $\alpha=1$.
To the zeroth order we get
\begin{align}
\overline{\langle\delta\rho[\br,\bu(\br)]\rangle}=\lim_{n\to 0}\langle\delta\rho[\br,\bu_1(\br)] \rangle_{\cH_0^n}=0,
\end{align}
since $\langle u^2(\br)\rangle_{\cH_0^n}=\infty$.  To capture the
physics correctly, we have to calculate mean FL density at least to 
first order in $v$ (see the discussion in
Sec.~\ref{sec:fridelirellevant}),
\begin{align}\label{eq:density average}
\langle\delta\rho[\br,\bu(\br)]\rangle=&-\beta\lim_{n\to 0}\sum_{\alpha=1}^n\langle\delta\rho[\br,\bu_1(\br)]\;\cH_{D}(\bu^{\alpha})\rangle_{\cH_0^n}\notag\\ &=\beta\lim_{n\to 0}\big\{\langle\delta\rho[\br,\bu_1(\br)]\;\cH_{D}(\bu_2)\rangle_{\cH_0^n}-\langle\delta\rho[\br,\bu_1(\br)]\;\cH_{D}(\bu_1)\rangle_{\cH_0^n}\big\}.
\end{align}
$u_2(\br)$ is the displacement field with replica index $\alpha=2$. First, we obtain the average of the long wavelength part of the FL density.
It can be shown that
\begin{align}
\langle\bnabla_{\bx}\bu^{\beta}(\br)\cos{[G_D \delta-\bG_D\bu^{\alpha}(\br_D)]}\rangle_{\cH_0^n}=\bnabla_{\bx}\Big\{& T\sin{(G_D\delta)}\;\langle\cos{[\bG_D\bu^{\alpha}(\br_D)]}\rangle_{\cH_0^n}\; \boldsymbol{\mathcal{G}}^{\alpha,\beta}(\br_D-\br)|\bG_D\rangle\notag\\& -T\cos{(G_D\delta)}\; \langle\sin{[\bG_D\bu^{\alpha}(\br_D)]}\rangle_{\cH_0^n} \boldsymbol{\mathcal{G}}^{\alpha, \beta}(\br_D-\br)|\bG_D\rangle\Big\},
\end{align}
where $\br=(\bx,z)$ and $\boldsymbol{\mathcal{G}}^{\alpha,
  \beta}(\br)$ is the propagator given by
Eq.~(\ref{eq:propagator}). Since
$\langle\sin{[\bG_D\bu^{\alpha}(\br_D)]}\rangle_{\cH_0^n}=
\langle\cos{[\bG_D\bu^{\alpha}(\br_D)]}\rangle_{\cH_0^n}=0$ this
contribution vanishes.

A finite difference between the expressions
$\langle\delta\rho(\bu_1)\;\cH_{D}(\bu_2)\rangle_{\cH_0^n}$ and
$\langle\delta\rho(\bu_1)\;\cH_{D}(\bu_1)\rangle_{\cH_0^n}$ appearing
in Eq.~(\ref{eq:density average}) can result from the thermal part of
the propagator that is diagonal in replica indices,
\begin{align}
\lim_{n\to 0}\big\langle\delta\rho[\bu^{\alpha}(\br)]\;\cH_{D}(\bu^{\beta})\big\rangle_{\cH_0^n} &=\lim_{n \to 0}\Big\{\frac{2 v_1\rho^2_0} {T}\int_{\br_D} \sum_{\bG\neq\mathbf{0}}e^{i\bG\cdot\bx}\Big\langle e^{-i\bG\cdot\bu^{\alpha}(\br)} \cos{\{G_D[\delta- u_{x}^{\beta}(\br_D)]\}}\Big\rangle_{\cH_0^n}\Big\}\notag\\&=\lim_{n \to 0} \Big\{\frac{v_1\rho^2_0}{T}\int_{\br_D} \sum_{\bG\neq\mathbf{0}}e^{i\bG\cdot\bx}\left( e^{iG_D\delta}I_++e^{-iG_D\delta}I_-\right)\Big\}\, ,
\end{align}
where $I_{\pm}=e^{-\frac{1}{2}\left\langle[\bG\cdot\bu^{\alpha}(\br) \pm \bG_D\cdot\bu^{\beta}(\br_D)]^2\right\rangle_{\cH_0^n}}$ and $\int_{\br_D}$ denotes the integration along the defect plane.
Analyzing $I_+$ ($I_-$), we conclude that it is nonzero only for $\bG=-\bG_D$ ($\bG=\bG_D$) and
\begin{align}
I_{\pm}\approx e^{\frac{G_D^2T}{4\pi c}\frac{\delta_{\alpha,\beta}}{|\br-\br_D|}}\left(\frac{L_a}{|\br-\br_D|}\right)^{2g}.
\end{align}
This yields
\begin{align}
\overline{\langle\delta\rho(\br)\rangle}&\approx\frac{v_1\rho^2_0}{T} \left(e^{-iG_D (x-\delta)} +e^{iG_D (x-\delta)} \right)\int_{\br_D}\left(\frac{L_a}{|\br-\br_D|}\right)^{2g}\left(e^{\frac{G_D^2T}{4\pi c}\frac{1}{|\br-\br_D|}}-1\right)\notag
\\&\approx \frac{4\pi v_1\rho^2_0L_a^2}{T}\cos{[G_D(x-\delta)]}\left(\frac{G_D^2T}{2\pi c}\right)^{2-2g}F\left(\frac{G_D^2 T}{4\pi c|x-\delta|}\right),
\end{align}
where $F(x)=\sum_{n=1}^{\infty}\frac{1}{n!}\frac{x^{2g+n-2}}{2g+n-2}$.
For very small temperatures the main contribution is
\begin{align}\label{eq:friedel irrelevantappendix}
\overline{\langle\delta\rho(\br)\rangle}&\approx\frac{v_1\rho^2_0G_D^2L_a}{c(2 g-1)} \cos{[G_D(x-\delta)]}\left(\frac{L_a}{|x-\delta|}\right)^{2 g-1}+\mathcal{O}(T).
\end{align}
The result captures the large scale behavior since it is valid on scales larger than $L\ge L_a$. Here $v_1$ denotes the effective defect strength measured on the scale $L_a$, and $|x-\delta|$ is the distance to the defect plane. Additional contributions to Eq.~(\ref{eq:friedel irrelevantappendix}), coming from the higher harmonics in $\cH_D$, are less important since they are proportional to the coefficients $v_k$ at scale $L=L_a$ and their amplitudes decay  as $|x-\delta|^{-2 k^2 g+1}$ with integer $k\geq 2$.

\section{Density oscillations in the presence of a relevant defect plane}
\label{app relevant}

In this appendix we study the displacement correlation functions and
average FL density profile for a relevant defect plane in the presence
of point impurities. As shown in the main text above, on sufficiently
large length scales pinning effects can be taken into account through
the boundary condition $u_x(\br_D)=0$ at the defect plane. For
simplicity we take the defect to be at the coordinate origin,
i.e., we set $\delta=0$.  First we calculate the generating function
\begin{align}
\overline{Z^n[\bj^{\alpha}(\br)]}=\int\mathcal{D}[\bu^{\alpha}]
e^{-\frac{\cH^n_0[\bu^{\alpha}]}{T}} e^{\sum_{\alpha}
\int_{\br}\bj^{\alpha}(\br)\bu^{\alpha}(\br)}
\prod_{\alpha,\br_D}\delta[u_{x}^{\alpha}(\br_D)].
\end{align}
Using the representation of the delta-function of Eq.~(\ref{eq:dirac
delta}) we get
\begin{align}
\overline{Z^n[\bj^{\alpha}(\br)]}=e^{\frac{T}{2}\sum_{\alpha,\beta}
\int_{\br_1,\br_2}\bj^{\alpha}(\br_1)\mathbf{\Gamma}^{\alpha,\beta}
(\br_1,\br_2)\bj^{\beta}(\br_2)},
\end{align}
where
\begin{align}
\mathbf{\Gamma}^{\alpha\beta}(\br_1,\br_2)=\boldsymbol{\mG}^{\alpha\beta}(\br_1-\br_2)
-\sum_{\gamma\kappa}\int_{\br_{D1}\br_{D2}}
\boldsymbol{\mG}^{\alpha\gamma}(\br_1-\br_{D1})|\hat{\bx}\rangle \;\widetilde{\boldsymbol{\mathcal{Q}}}^{-1}_{\gamma\kappa}
(\br_{D1}-\br_{D2})
\langle\hat{\bx}|\boldsymbol{\mG}^{\kappa\beta}(\br_2-\br_{D2})
\end{align}
and $\widetilde{\boldsymbol{\mathcal{Q}}}$ is given by Eq.~(\ref{eq:Q}) and ${\boldsymbol{\mG}}^{\alpha,\beta}$ is the inverse of ${\boldsymbol{\mG}}^{-1}_{\alpha,\beta}$ given by Eq.~(\ref{eq:propagator}).
The displacement correlation function is given by the relation $\overline{\langle
\bu(\br)\bu(\br)\rangle}=\lim_{n\to0}\frac{\delta^2 \overline{Z^n}
}{\delta\bj^{1}(\br)\delta\bj^{1}(\br)}\Big|_{j^{\alpha}=0}$.
Denoting by $\bq$ the in-plane momentum, we get in momentum space
\begin{align}\label{eq:correlation pinned Fourier2}
\widetilde{\mG}_{pin,ij}(x,x;\bq)=\lim_{n\to 0}
 \left[  \widetilde{\boldsymbol{\mG}}_{ij}^{11}(0,\bq)-
\sum_{\alpha\gamma} \widetilde{\boldsymbol{\mG}}^{1\alpha}_{ix}(|x|,\bq)\widetilde{\boldsymbol{\mathcal{Q}}}^{-1} _{\alpha\gamma}(\bq)\widetilde{\boldsymbol{\mG}}^{1\gamma}_{xj}(|x|,-\bq)\right],
\end{align}
where $\mG_{pin,ij}(x,x';\br_{\parallel}-\br'_{\parallel})=T^{-1}\overline{\langle
  u_{i}(\br)u_{j}(\br')\rangle}$ with $\br=(x,\br_{\parallel})$. The indices $i,j$ take the values
$x,y$. All
propagators and their inverse that appear in the previous equations
have the form $\boldsymbol{X}_{\alpha,\beta}=\delta_{\alpha,\beta}
\boldsymbol{X}_d+\boldsymbol{X}_n$. The only nonzero contribution to
the second term of Eq.~(\ref{eq:correlation pinned Fourier2}) comes
from the product of all ''diagonal'' parts ($\boldsymbol{X}_d$) of the propagators
 or only one ''nondiagonal'' ($\boldsymbol{X}_n$) and
two diagonal (in replica indices) in the limit $n\rightarrow
0$. Denoting by $X_{ij,a}=\lim_{n\to 0}\langle
\hat{i}|\boldsymbol{X}_{a}|\hat{j}\rangle$, where $a=d,n$, one has
\begin{align}\label{eq:G_{pin}}
\widetilde{\mG}_{pin,ij}(x,x;\bq)&=\widetilde{\mG}_{ij,d}(0,\bq)-\widetilde{\mG}_{ix,d}(|x|,\bq)\;\widetilde{\mG}_{jx,d}(|x|,\bq)\;\widetilde{{\mathcal{Q}}}^{-1}_{d}(\bq)
\notag\\& +\widetilde{\mG}_{ij,n}(0,\bq)-\widetilde{\mG}_{ix,d}(|x|,\bq)\;\widetilde{\mG}_{jx,d}(|x|,\bq)\;\widetilde{{\mathcal{Q}}}^{-1}_{n}(\bq) +
\widetilde{{\mathcal{Q}}}^{-1}_{d}(\bq) \; \widetilde{\mG}_{jx,n}(|x|,\bq)\; \widetilde{\mG}_{ix,d}(|x|,\bq)\notag\\ &+ \widetilde{{\mathcal{Q}}}^{-1}_{d}(\bq) \; \widetilde{\mG}_{ix,n}(|x|,\bq)\; \widetilde{\mG}_{jx,d}(|x|,\bq).
\end{align}
For isotropic elasticity the relation
\begin{align}
\widetilde{\mG}_{ix,d}(|x|,\bq)\;\widetilde{\mG}_{jx,d}(|x|,\bq)\;\widetilde{{\mathcal{Q}}}^{-1}_{d}(\bq)=(\hat{\bx}\cdot \hat{i})(\hat{\bx}\cdot \hat{j})
\widetilde{\mG}_{ij,d}(2|x|,\bq)
\end{align}
holds.  After integrating the terms of the second and the third line
of Eq.~(\ref{eq:G_{pin}}) over $\bq$ we find that the displacement
correlations on scales larger than $L_v$ (\ref{eq:Lv}) read
\begin{align}
\overline{\langle u_i(\br) u_j(\br)\rangle}&=\lim_{n\to 0}\left\{T\boldsymbol{\mG}_{ij}^{11}(0,\mathbf{0})-T(\hat{\bx}\cdot\hat{i})(\hat{\bx}\cdot\hat{j})\boldsymbol{\mG}_{xx}^{11}(2|x|,\mathbf{0})\right\}.
\end{align}

Next, we shall calculate the disorder and thermal average of the FL density
\begin{align}\label{eq:expectation ro}
\overline{\langle\delta\rho(\br)\rangle}=\rho_0\sum_{\mathbf{G}\neq 0}e^{i\bG\bx}\overline{\langle
e^{-i\bG\bu}\rangle} \, ,
\end{align}
using
\begin{align}
\overline{\langle
e^{-i\bG\bu}\rangle}=\lim_{n\to 0}\langle
e^{-i\bG\bu_1}\rangle=\lim_{n\to 0} e^{-\frac{1}{2}\langle{(\bG\bu_1(\br))^2}\rangle}
\end{align}
and
\begin{align}
\lim_{n\to 0}\langle{[\bG\bu_1(\br)]^2\rangle}=\lim_{n\to 0}T\left[G^2\; \boldsymbol{\mG}_{xx}^{11}(\mathbf{0})-(\bG\cdot\hat{\bx})^2\boldsymbol{\mG}_{xx}^{11}(2|x|,\mathbf{0})\right].
\end{align}
Since $\lim_{n\to 0}\boldsymbol{\mG}_{xx}^{11}(\mathbf{0})$ is
divergent we conclude that only terms with a reciprocal vector $\bG$
perpendicular to the defect plane contribute in Eq.~(\ref{eq:expectation
  ro}) and
\begin{align}
\overline{\langle\delta\rho(\br)\rangle}=2\rho_0\sum_{m>0}\cos{(mG_D x)} \left(\frac{L_v}{|x|}\right)^{m^2g},
\end{align}
\end{widetext}where $m$ is an integer.

\section{Sample-to-sample fluctuations of the magnetic susceptibility}
\label{app2}
In this appendix we examine the influence of planar defects on the
longitudinal magnetic susceptibility.  An infinitesimal change in the
longitudinal magnetic field $\delta H_z\mathbf{\hat z}$ changes the
Hamiltonian of Eq.~(\ref{eq:Ham}) in the case of an uniaxial displacement
field perpendicular to the defects by
\begin{align}\label{eq:new}
 \delta \cH=-\frac{\phi_0\rho_0}{4\pi}
\int d^3r\;
 \delta H_z\partial_{x}u.
\end{align}
Since the change of the magnetic induction is
$B=\rho_0\phi_0\partial_x u$, the longitudinal magnetic susceptibility
reads
\begin{align}
\chi=\rho_0\phi_0\frac{\partial\langle\partial_{x} u \rangle}{\partial \delta H_z}=-\frac{4\pi}{V}\frac{\partial^2 {F}}{\partial \delta H_z^2}
\end{align}
where $F$ is the free energy. It is convenient to consider a
generalization of this model to $d$ dimensions where $\bx$ is a
$d-2$-dimensional vector and $x_1$ is the component of $\bx$ in the
direction of the displacement $u$. Applying the transformation $u\to
u+h x_1/c_{11}$, the additional term given by Eq.~(\ref{eq:new}) can
be shifted away yielding
\begin{align}\label{eq:newH}
\cH(h,u)=&\cH_0(u)-\frac{h^2}{2c_{11}}V+\cH_D\left(u+h x_1/c_{11}\right)\, ,
\end{align}
where $V=L_x^{d-2}L_zL_y$ and $h=\delta H_z{\rho_0\phi_0}/(4\pi)$. The pinning energy of planar defects $\cH_D$ can be written as
\begin{align}
\cH_D(u)=&\int d\mathbf{r} V_D(\bx)\rho_s(u,\br) +\int d\mathbf{r} V_D(\bx)\rho_p(u,\br)\notag\\=&\cH_D^s+\cH_D^p,
\end{align}
where $\rho_s$ and $\rho_p$ are the slowly varying and periodic part
of the FL density, respectively.  Next, we would like to compute the
average $\overline{\chi}$. The free energy is given by
\begin{align}
F(h)=&-T \log{\left(\int \mathcal{D}u e^{-\cH(h,u)/T}\right)}\notag\\
=&-\frac{h^2}{2 c_{11}}V-\rho_0 \frac{h}{c_{11}}\int d\mathbf{r}{V_D(\bx)}-T\log{Z_1},
\end{align}
where $Z_1$ is the partition function for
$\cH_1(h,u)=\cH_0(u)+\cH_D^s(u)+\cH_{D}^p\left(u+h
  x_1/c_{11}\right)$. Using replicas, the disorder averaged free
energy can be written as
\begin{align}\label{eq:averagefree}
\overline{F}=-\frac{h^2}{2 c_{11}}V- T\lim_{n\to 0}\frac{\overline{Z_1^n}-1}{n}. 
\end{align}
Here $\overline{Z_1^n}=\int \mathcal{D}[u^{\alpha}] \exp{[-\cH_1^n/T]}$ where $\cH_1^n$ is the replica Hamiltonian that follows from $\cH_1(h,u)$. Since $\cH_1(h,u)$ has the same statistical properties as $\cH(0,u)$, i.e., it yields the same replica Hamiltonian, the only dependence on $h$ in $\overline{F}$ comes from the first quadratic term in Eq.~(\ref{eq:averagefree}). Due to this so-called statistical tilt symmetry \cite{Schulz+88}, the disorder averaged susceptibility
\begin{align}
 \overline{\chi}=-\frac{4\pi}{V}\frac{\partial^2 \overline{F}}{\partial \delta H_z^2}=\frac{(\rho_0\phi_0)^2}{4\pi c_{11}}
\end{align}
is disorder independent. The important quantity are the
sample-to-sample variations of the susceptibility. The free energy can
be written as
\begin{align}
F(h)=-\frac{h^2}{2c_{11}}V+F_0-T\log{\langle e^{-\cH_{D}\left(u+{h x_1}/{c_{11}}\right)/T} \rangle_{\cH_0}},
\end{align}
where $F_0$ is the free energy of the system that is described by the
elastic Hamiltonian only. To first order in perturbation theory with
respect to $\sim v$ we get
\begin{align}
\Delta F(h)=F(h)-\overline{F(h)}=\left\langle \cH_D\left(u+{h x_1}/{c_{11}}\right)\right\rangle_{\cH_0}.
\end{align}
For a system of linear size $L_x$ in the $x$ direction we find
\begin{widetext}\begin{align}
\overline{\Delta F(h_{1})\Delta F(h_{2})}=2{(v\rho_0)^2}{n_{pd}}(L_yL_z)^2\int \dif\mathbf{x} \sum_{n>0}^{[\ell/\xi]_G}
e^{-(nG_D)^2\langle u^2 \rangle_{\cH_0}}  \cos{\left[nG_D\frac{\left(h_{1}-h_{2}\right)x_1}{c_{11}}\right]},
\end{align}\end{widetext}
where $n_{pd}$ denotes the density of defects. Differentiation with
respect to $h_1$ and $h_2$ leads to
\begin{align} \frac{\overline{\Delta\chi^2}}{\overline{\chi}^2}=\frac{R_D''''(0)L_x^{\epsilon}}{5c_{11}^2}\sim
\left(\frac{L_x}{L_D}\right)^{\epsilon} \, ,
\end{align}
where we have taken into account the irrelevance of thermal
fluctuation.  Since $\epsilon >0$ for $d<6$, the sample-to-sample
fluctuations grow with the scale $L_x$. One cannot expect that this
result is quantitatively correct for large $L_x$. However,
qualitatively it demonstrates the relevance of defects and it is a
signature of a glassy phase. For $L_x>L_D$ we expect that
${\overline{(\Delta\chi)^2}}/{\overline{\chi}^2}$ approaches a finite
universal value for $d<6$.

\section{Positional correlation function for $6>d>4$}
\label{app3}

In this appendix the positional correlation function of the FLL with
planar defects will be calculated, using  perturbation theory and
results from the functional RG analysis presented in
Sec.~\ref{section:weak}. The positional correlation functions have
been calculated before for the FLL with point impurities for an uniaxial
displacement field \cite{Giamarchi+94,Giamarchi+95} and for a vector
displacement field \cite{Emig+99,bogner+01}. We perform the
computations along the lines of these references.

In a functional RG procedure, after integrating out fast modes in an
infinitesimal shell with $\Lambda/b\leq q\leq \Lambda$, one can choose
to keep the cutoff in momentum space fixed using the rescaling
\begin{align}
\bx'&=\frac{\bx}{b}\qquad
\bz'=\frac{\bz}{b^{\chi}}\notag\\
\bq_x'&=b \bq_x\quad
\bq_z'=b^{\chi}\bq_z,
\end{align}
where $\bz=(y,z)$. The displacement field is not rescaled due to the
periodicity of $R_D$. This implies $u(\bq)=b^{d-2+2\chi}u'(\bq')$.  We
need to obtain the RG flow of the correlation function
\begin{align}\label{eq:rescaledcorr}
&\langle u(\bq_1)u(\bq_2)\rangle=Z^{-1}\int \mathcal{D}u(\bq)e^{-\beta \cH} u(\bq_1)u(\bq_2)\notag\\
&=Z^{-1} \int \mathcal{D}u^{<}(\bq)u(\bq_1)u(\bq_2)\int \mathcal{D}u^{>}(\bq)e^{-\beta \cH}\notag\\
&=Z^{-1}\int \mathcal{D}u^{<}(\bq)u(\bq_1)u(\bq_2)e^{-\beta \cH_l(u^{<}(\bq))}\notag\\
&=b^{2(d-2+2\chi)}\langle u'(\bq'_1)u'(\bq'_2)\rangle \, ,
\end{align}
where $\cH=\cH_0+\int V_D(x)\rho(u,\br)$. Here $u^{<}(\bq)$ are modes
that satisfy $q<\Lambda/b$ and $\cH_l$ is the Hamiltonian that
applies to the scale $l=\log b$ with $b$ very close to unity. Using
Eq.~(\ref{eq:rescaledcorr}), we obtain a differential equation for
$\langle u(\bq_1)u(\bq_2)\rangle$. We get
\begin{align}\label{eq:rescaledcorr1}
\langle u(\bq_1)u(\bq_2)\rangle= b^{2(d-2+2\chi)}\langle u'(\bq'_1)u'(\bq'_2)\rangle
\end{align}
where the only restriction on $b$ is $q_{i}<\Lambda/b$, $i=1,2$.

\begin{widetext}
  First, we calculate $\langle u(\bq_1)u(\bq_2)\rangle$ to lowest
  order in $v$,
\begin{align}
&\langle u(\bq_1)u(\bq_2)\rangle= \langle u(\bq_1)u(\bq_2)\rangle_{\cH_0}+\frac{1}{2T^2}\lim_{n\to 0}\sum_{\beta,\gamma}\int_{\bx,\bz_1,\bz_2}\left \langle u_{1}(\bq_1)u_{1}(\bq_2)R_D[u^{\beta}(\bx,\bz_1)-u^{\gamma}(\bx,\bz_2)] \right\rangle_{\sum_{\alpha}\cH_0(u^{\alpha})},
\end{align}
where $u_1$ is the displacement field with replica index $\alpha=1$.
We use the  periodicity of $R_D$ by writing $R_D(u)=\sum_{n}R_n \cos{(n G_D u)}$. From this we find that \emph{at the planar glass fixed point}
\begin{align}\label{eq:corrfixed}
\langle u(\bq_1)u(\bq_2)\rangle=\frac{-(2\pi)^{d+2}}{(c_{11}q_{1x}^2)^2}R_D^{*''}(0)\delta(\bq_{1x}+\bq_{2x})\delta(\bq_{1z})\delta(\bq_{2z}),
\end{align}
where the rescaled fixed point correlator is
$R_{D}^{*''}(u)=-\frac{\epsilon c_{11}^2\Lambda^{\epsilon}}
{6K_d}\left[(u-\ell/2)^2-\ell^2/12\right]$ for $0\le u<\ell$.
Choosing $b=\Lambda/\max\{q_{1x},q_{2x}\}$ in
Eq.~(\ref{eq:rescaledcorr1}) so that it is justified to calculate the
correlation function appearing on the right hand side of
Eq.~(\ref{eq:rescaledcorr1}) at the fixed point, and using the result
of Eq.~(\ref{eq:corrfixed}), we get
\begin{align}\label{eq:corr}
\langle u(\bq_1)u(\bq_2)\rangle=\frac{(2\pi)^{2d}}{36 S_{d-2}}\frac{\Lambda^{2(6-d)}}{q_{1x}^{d-2}}\epsilon \ell^2 \delta(\bq_{1x}+\bq_{2x})\delta(\bq_{1z})\delta(\bq_{2z}).
\end{align}
Note that in Eq.~(\ref{eq:corr}) only displacements with
$\bq_{1x}=-\bq_{2x}$ are coupled as it would be the case for a
quadratic Hamiltonian. Therefore one can write down an effective quadratic
Hamiltonian in the $\bq_x$-momentum space that reproduces the
correlations to 
first order in $\epsilon$. The positional correlation
function shows the power low behavior
\begin{align}
S_{G_D}\approx|\bx|^{-\epsilon\left(\pi/3\right)^2}.
\end{align}
We point out that this result is valid only for $d>4$. In $d\le 4$
dimensions the part of the pinning potential related to the slowly
varying part of the FL density also becomes relevant and further
analysis is needed, see Sec.~\ref{section:weak}.

\section{List of recurrent symbols}
\begin{tabular*}{0.95\textwidth}{@{\extracolsep{\fill}} l  l  l  }
\hline\hline
Symbol & Quantity & Definition \\ \hline 			

$a$ & flux line lattice constant &  \\

$\mathbf{B}$ & magnetic induction & \\

$c_{ii}\quad i=1,4,6$ & elastic constants & Eq.~(\ref{eq:elastic Ham})\\

$\bG$ & reciprocal lattice vector & \\

$g=\frac{3}{8}\eta(\frac{a}{\ell})^2$ & parameter controlling the relevance of the single defect plane & Eq.~(\ref{eq:flow})\\

$G_D=\frac{2\pi}{\ell}$& shortest reciprocal lattice vector perpendicular to the defect(s)    &  Sec.~\ref{section:single} \\

$\cH_0$ & elastic energy of distortions of the FLL & Eq.~(\ref{eq:elastic Ham})\\

$\cH_{P}$ & pinning energy of point impurities & Eq.~(\ref{eq:pointpinning})\\

$\cH_0^n$ & effective quadratic replica Hamiltonian for defect free system & Eq.~(\ref{eq:Replica Hamiltonian quadratic}) \\

$\cH_D$ & pinning energy of planar defect & Eq.~(\ref{eq:defect H})\\

$J$ & current density & \\

$L_{\xi}$ & Larkin length & Eq.~(\ref{eq: Larkinlength})\\

$L_{a}$ & positional correlation length & Sec.~\ref{section:BG}\\

$\ell_D$& mean distance between defects& Sec.~\ref{section:weak}\\

$L_D$ & Larkin length for planar defects & Sec.~\ref{section:weak}\\

$n_{imp}$ & density of point impurities   & \\

$n_D$ & unit vector perpendicular to the defect plane   &  \\

$\br_D$&  position vector of the defect plane  &  Eq.~(\ref{eq:defect-para}) \\

$R_P$ & point disorder correlation function & Eq.~(\ref{eq:R(u)})\\

$R_D$ & planar disorder correlation function & Eq.~(\ref{eq:RD})\\

$R_D^*$& fixed point value of $R_D$ & \\

$S_{\bG}(\br)$ & positional correlation function& Sec.~\ref{section:BG}\\

$S_{d}$ & surface of $d$ dimensional unit sphere & \\

$T$ & temperature & \\

$u_i(y,z)=u(x_i,y,z)$& displacement field at the planar defect with $x=x_i$& \\

$u_x$& displacement field perpendicular to the defects& \\

$v_p$ & strength of point impurities   &  Sec.~(\ref{section:BG}) \\

$V_P$ & pinning potential resulting from point impurities & Eq.~(\ref{section:BG})\\

$v$ &  defect strength  &   Sec.~\ref{section:single}\\

$V_D$ & pinning potential resulting from planar defects &  Sec.~\ref{section:weak}\\

$[x]_G$ & the closest integer to x & \\

$\delta$ & defect distance to the origin   &  \\

$\zeta$& roughness coefficient &Sec.~\ref{section:BG} \\

$\eta$ & power law exponent of positional correlation function in the Bragg glass regime & Sec.~\ref{section:BG}\\

$\Lambda$ & momentum cutoff & Sec.~\ref{section:BG}\\

$\lambda$ & penetration depth & Sec.~\ref{section:BG} \\

$\xi_{RF}$&  roughness exponent in random force regime & Sec.~\ref{section:BG}\\

$\xi_{RM}$&  roughness exponent in random manifold regime & Sec.~\ref{section:BG} \\

$\xi_{BG}$&  roughness exponent in the Bragg glass regime & Sec.~\ref{section:BG} \\

$\xi$ & superconductor coherence length& \\

$\xi_c$& correlation length& \\

$\rho(\bu,\br)$ & flux line density & Eq.~(\ref{eq:density FLs})\\

$\rho_0$ & background FL density & Eq.~(\ref{eq:density FLs}) \\

$\Sigma_{y(z)}$ & interface tension of domain wall parallel to $\textbf{x}$ and $z$ ($y$) axes & Eq.~(\ref{eq:newterm}) \\

$\phi_0$& flux quantum & Sec.~\ref{section:BG}\\
\hline\hline
\end{tabular*}

\end{widetext}

\end{document}